\pdfoutput=1

\documentclass[12pt,preprint]{aastex}

\usepackage{color}

\usepackage{amsmath,amsthm}

\shorttitle{On the Helicity of Open Magnetic Fields}
\shortauthors{Prior \& Yeates}

\newcommand{\pder}[2]{\frac{\partial#1}{\partial#2}}
\newcommand{\rmd}{\mathrm{d}}
\newcommand{\Hw}{H^{\rm W}}

\begin{document}


\title{On the Helicity of Open Magnetic Fields}


\author{C. Prior, A.R. Yeates}
\affil{Department of Mathematical Sciences, Durham University, Durham, DH1 3LE, UK}
\email{anthony.yeates@durham.ac.uk}


\begin{abstract}
We reconsider the topological interpretation of magnetic helicity for magnetic fields in open domains, and relate this to the relative helicity. Specifically, our domains stretch between two parallel planes, and each of these ends may be magnetically open. It is demonstrated that, while the magnetic helicity is gauge-dependent, its value in any gauge may be physically interpreted as the average winding number among all pairs of field lines with respect to some orthonormal frame field. In fact, the choice of gauge is equivalent to the choice of reference field in the relative helicity, meaning that the magnetic helicity is no less physically meaningful. We prove that a particular gauge always measures the winding with respect to a fixed frame, and propose that this is normally the best choice. For periodic fields, this choice is equivalent to measuring relative helicity with respect to a potential reference field. But for aperiodic fields, we show that the potential field can be twisted. We prove by construction that there always exists a possible untwisted reference field.
\end{abstract}




\section{Introduction}

Magnetic helicity $H({\bf B})=\int_V{\bf A}\cdot{\bf B}\,\rmd^3x$ has long been recognized as an important dynamical invariant in ideal magnetohydrodynamics, with applications ranging from laboratory plasmas to astrophysical objects \citep{Brown1999a}. Here ${\bf A}$ is a vector potential for the magnetic field ${\bf B}=\nabla\times{\bf A}$, and it is a fundamental property of $H({\bf B})$ that the integral is independent of the particular gauge chosen for ${\bf A}$, provided that $V$ is simply connected and magnetically closed ($B_n=0$ on the boundary $\partial V$). Analogous invariants exist for other solenoidal vector fields, notably the vorticity in fluid mechanics \citep{Moffatt1969a}.

Physically, $H({\bf B})$ may be interpreted as a measure of the average topological linking of the magnetic field lines of ${\bf B}$ \citep{Moffatt1969a,Arnold1986,Arnold1998}. One way to see this is to consider a special magnetic configuration where ${\bf B}$ is confined to two (or more) linked magnetic flux tubes that are closed and untwisted \citep[see, for example,][]{Moffatt1992a}. Another way is to write ${\bf A}$ in Coulomb gauge ($\nabla\cdot{\bf A}=0$), whence, providing that $B_n=0$ on the whole boundary of $V$, it has the expression
\begin{equation}
{\bf A}({\bf x}) = \frac{1}{4\pi}\int_V \frac{{\bf B}({\bf y})\times{\bf r}}{|{\bf r}|^3}\,\rmd^3y,
\label{eqn:bs3}
\end{equation}
where ${\bf r}={\bf x}-{\bf y}$ \citep{Cantarella2001}. It follows that $H({\bf B})$ may be written as
\begin{equation}
H({\bf B})=\frac{1}{4\pi}\int_V\int_V{\bf B}({\bf x})\cdot\frac{{\bf B}({\bf y})\times{\bf r}}{|{\bf r}|^3}\,\rmd^3x\,\rmd^3y.
\label{eqn:hlink}
\end{equation}
This is the flux-weighted average, over all pairs of magnetic field lines $\rmd{\bf x}/\rmd s={\bf B}(\bf {x})$, $\rmd{\bf y}/\rmd s={\bf B}(\bf {y})$, of the Gauss linking integral
\begin{equation}
\label{gl}
L({\bf x},{\bf y}) = \frac{1}{4\pi}\oint_{{\bf x}(s)}\oint_{{\bf y}(s)}\frac{d{\bf x}}{ds}\cdot\frac{d{\bf y}}{ds}\times\frac{\bf r}{|{\bf r}|^3}\,\rmd s\,\rmd s'.
\end{equation}
The Gauss integral is integer-valued and measures the net linking of a pair of closed curves \citep{Ricca2011}.

Unfortunately, the gauge invariance of $H$ relies on the condition $B_n|_{\partial V}=0$. In astrophysical situations such as the solar atmosphere, this condition is generally violated. In a seminal paper, \citet{Berger1984ay} showed how gauge invariance may be restored by measuring the helicity with respect to a chosen reference magnetic field ${\bf B}'$ sharing the same distribution of $B_n$ on $\partial V$. This relative helicity, which we shall denote $H_{{\bf B}'}({\bf B})$, is then an ideal invariant under motions that vanish on $\partial V$. It has since been widely applied to the open magnetic fields arising in solar physics \citep[see the review by][]{Demoulin2007c}.

This work is motivated by a fundamental question: is there a topological interpretation of relative helicity in open fields analogous to the linking number interpretation of $H$ (Equation \ref{eqn:hlink}) in closed fields? Since the magnetic field lines are no longer closed curves, they no longer have invariant Gauss linking integrals. However, one can construct alternative invariants for pairs of curves stretching between two planes, provided that the end-points are held fixed \citep{Berger1986,Berger1993b}. And indeed we will show in Section \ref{sec:wind} that it is possible to express both $H$ and $H_{{\bf B}'}$ in terms of these ``winding numbers''. The fact that there are multiple ways of defining such invariant winding numbers reflects the fact that neither $H$ nor $H_{{\bf B}'}$ is uniquely defined for an open field. Rather, $H$ depends on the choice of gauge, and $H_{{\bf B}'}$ on the choice of reference field. In fact, we argue in Section \ref{sec:hr} that $H$ is no less meaningful than $H_{{\bf B}'}$ in an open field, despite the fact that the latter has been used preferentially in applications.

In solar physics, the non-uniqueness of $H_{{\bf B}'}$ has almost universally been circumvented by choosing ${\bf B}'$ to be the unique potential field ${\bf B}^{\rm p}$ matching $B_n$ on the boundary of the domain. The potential field is well-defined and has the minimum magnetic energy of all fields matching the same boundary conditions. In the case of magnetic fields rooted in a single planar boundary, $H_{{\bf B}^{\rm p}}$ has been shown explicitly to be an average winding number \citep{Berger1986,Demoulin2006}. This physical interpretation has been used to infer the injection of relative helicity into the Sun's corona by tracking the winding of magnetic field lines by their footpoint motions on the photospheric boundary \citep{Demoulin2007c}. However, there are two limitations that prevent $H_{{\bf B}^{\rm p}}$ from being a perfect helicity measure. The first limitation is that, if the boundary conditions $B_n|_{\partial V}$ are changing in time, then the reference field ${\bf B}^{\rm p}$ will itself change in time, and usually in a non-ideal way. This means that the evolution of the relative helicity will mix up both real topological changes in ${\bf B}$ and those simply due to the change of ${\bf B}^{\rm p}$. The second limitation is that, in a domain with more than one boundary where $B_n\neq 0$, the interpretation of $H_{{\bf B}^{\rm p}}$ as measuring the average winding number breaks down. This is shown in Section \ref{sec:hr}. Our central idea in this paper is that these limitations may be overcome by defining helicity not through $H_{{\bf B}^{\rm p}}$, but by fixing a special gauge in $H$. Fixing the gauge of $H$ will always create an ideal invariant that is (trivially) gauge independent. Our main contribution is to show in Section \ref{sec:bs2} how this invariant is physically meaningful.

It should be mentioned that several authors have proposed other alternatives to the widely used $H_{{\bf B}^{\rm p}}$. For example, \citet{Longcope2008} have explored different choices of reference field for relative helicity in sub-volumes of the solar corona. \citet{Low2006f} has proposed a ``Lagrangian helicity'' that decomposes ${\bf B}$ at some initial time into a toroidal and a poloidal component, then measures the linking between the two components mapped back to the initial configuration \citep[see also][]{Webb2010a,Low2011h}. This retains a freedom in the choice of the initial toroidal-poloidal decomposition. Closer in spirit to the present paper, \citet{Hornig2006} proposes to define $H$ completely with a particular choice of gauge, namely $\nabla^\perp\cdot{\bf A}=0$ on the boundary (where $\nabla^\perp$ denotes the component of the gradient tangential to the boundary). \citet{Jensen1984w} also imposed a gauge condition - that ${\bf n}\times{\bf A}={\bf n}\times{\bf A}^{\rm p}$ on $\partial V$ - to uniquely define their version of relative helicity, which has the form
\begin{equation}
H^{\rm JC}=\int_V{\bf A}\cdot{\bf B}\,\rmd^3x - \int_V{\bf A}^{\rm p}\cdot{\bf B}^{\rm p}\,\rmd^3x.
\end{equation}
Such gauge conditions are also frequently used to simplify the calculation of $H_{{\bf B}^{\rm p}}$ in practice \citep{Demoulin2007c}. For the particular case of a cylindrical domain, \citet{Low2011h} introduced an ``absolute helicity'' that is similarly based on fixing a particular gauge of ${\bf A}$ (related to a toroidal-poloidal, or Chandrasekhar-Kendall decomposition). The geometric characterisation of the helcity in any gauge which we highlight in this study allows for comparison of our fixed gauge measue with these alternatives. It is demonstrated in Section  \ref{sec:gen} that all choices expect the one we propose in this paper measure the field-line winding in a manner which is not wholly physically meaningful.

The layout of this paper is as follows. We briefly review the standard definitions of $H$ and of the relative helicity $H_{{\bf B}'}$ in Section \ref{sec:prelim}, before introducing an important special gauge in Section \ref{sec:bs2} for fields in a cylinder, which we call the ``winding gauge''. Section \ref{sec:wind} then presents the main contributions of this paper: (i) that $H$ is physically meaningful in any gauge, and (ii) that the winding gauge best captures our intuitive idea of field line winding. In Section \ref{sec:hr} we investigate how the ``winding'' helicity relates to the relative helicity. Conclusions are summarized in Section \ref{sec:conclusions}.

\section{Preliminaries} \label{sec:prelim}

\begin{figure}[ht]
\begin{center}
\includegraphics[width=14cm]{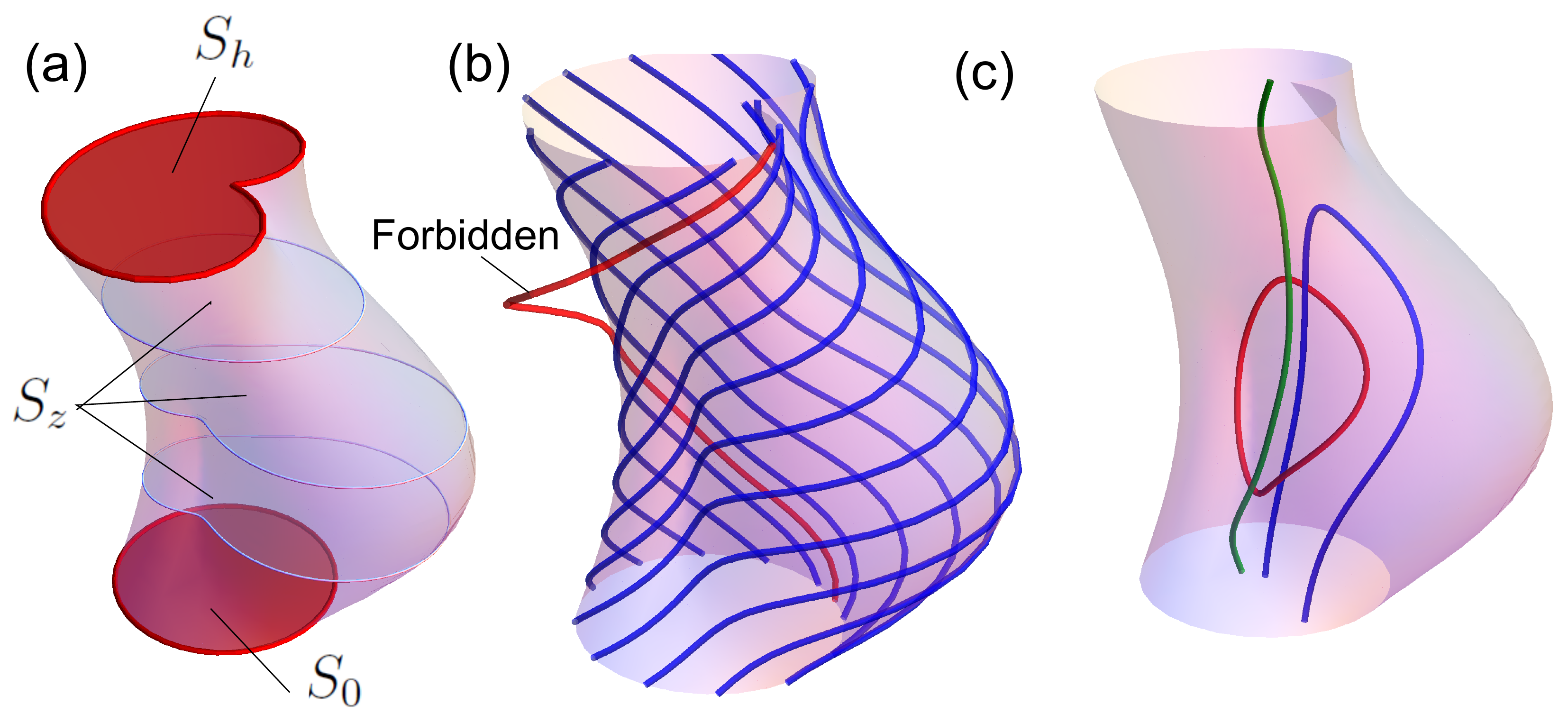}
\caption{The domain $V$ and its admissible magnetic field lines. Panel (a) depicts an example domain. Panel (b) shows (in  blue) a set of admissible field lines which are tangent to the boundary, and a red forbidden field line that is not allowed in this Paper. Panel (c) shows the three possible connectivities that an admissible field line may have.}
\label{fig:defs}
\end{center}
\end{figure}

Throughout, we shall consider magnetic fields on a domain $V\in \mathbb{R}^3$ where $V=S_z\times[0,h]$ for a set of simply-connected regions $S_z\subset\mathbb{R}^2$, $z\in[0,h]$, whose boundaries $\partial S_z$ vary continuously with $z$. An example is shown in  Figure \ref{fig:defs}(a). Each of the foliating surfaces $S_z$ has the same normal vector ${\bf \hat{z}}$. The boundary of $V$ consists of the lower boundary surface $S_0$, the upper boundary surface $S_h$, and the set $S_s = \left\{\partial S_{z}\vert z\in(0,h)\right\}$, \textit{i.e.} $\partial V = S_0\cup S_s\cup S_h$. We define a Cartesian co-ordinate system $\{\hat{\bf {e}}_1,\hat{\bf{e}}_2,\hat{\bf{z}}\}$ for $V$ with the pair $\{\hat{\bf{e}}_1,\hat{\bf{e}}_2\}$ spanning the surfaces $S_z$.

We consider magnetic fields ${\bf B}$ which are either tangent or zero on the side boundary $S_s$, but place no restrictions on the end boundaries $S_0$ and $S_h$. The conditions on $S_s$ forbid magnetic field lines from leaving the boundary (see Figure \ref{fig:defs}b). If $B_z$ has the same sign everywhere in $V$, then the magnetic field will essentially be a directional flow through the domain, akin to a magnetic flux rope. But our set of admissible magnetic fields is wider and allows for a mixture of field lines linking the two end planes, field lines that are looped, and those that are closed (examples are depicted in  Figure \ref{fig:defs}c). It also allows for fields in a half-space $h\rightarrow \infty,\,\,S_z = \mathbb{R}^2,\forall z \in[0,\infty)$. In this case a reasonable definition of helicity requires that the field decays to zero towards infinity, implying a looped field of the type discussed by \cite{Demoulin2006c}.

\subsection{Magnetic Helicity}

We shall denote the magnetic helicity by
\begin{equation}
H({\bf B}) = \int_V{\bf A}\cdot{\bf B}\,\rmd^3x.
\end{equation}
In a magnetically-open domain $V$, the helicity depends on the gauge of ${\bf A}$. For under a gauge transformation ${\bf A} \rightarrow {\bf A}' = {\bf A} + \nabla\chi$, we find
\begin{equation}
H \rightarrow H'=H + \oint_{\partial V}\chi B_n\,\rmd^2x.
\end{equation}
So if the normal magnetic field $B_n$ is non-zero anywhere on the boundary $\partial V$, we can change $H$ by changing the gauge $\xi$. For our domain (described above) we have $B_n=0$ on the side boundary $S_s$, so
\begin{equation}
H \rightarrow H'=H + \int_{S_h}\chi B_z\,\rmd^2x - \int_{S_0}\chi B_z\,\rmd^2x.
\label{eqn:hgauge}
\end{equation}

\subsection{Relative Helicity}

\begin{figure}[ht]
\begin{center}
\includegraphics[width=10cm]{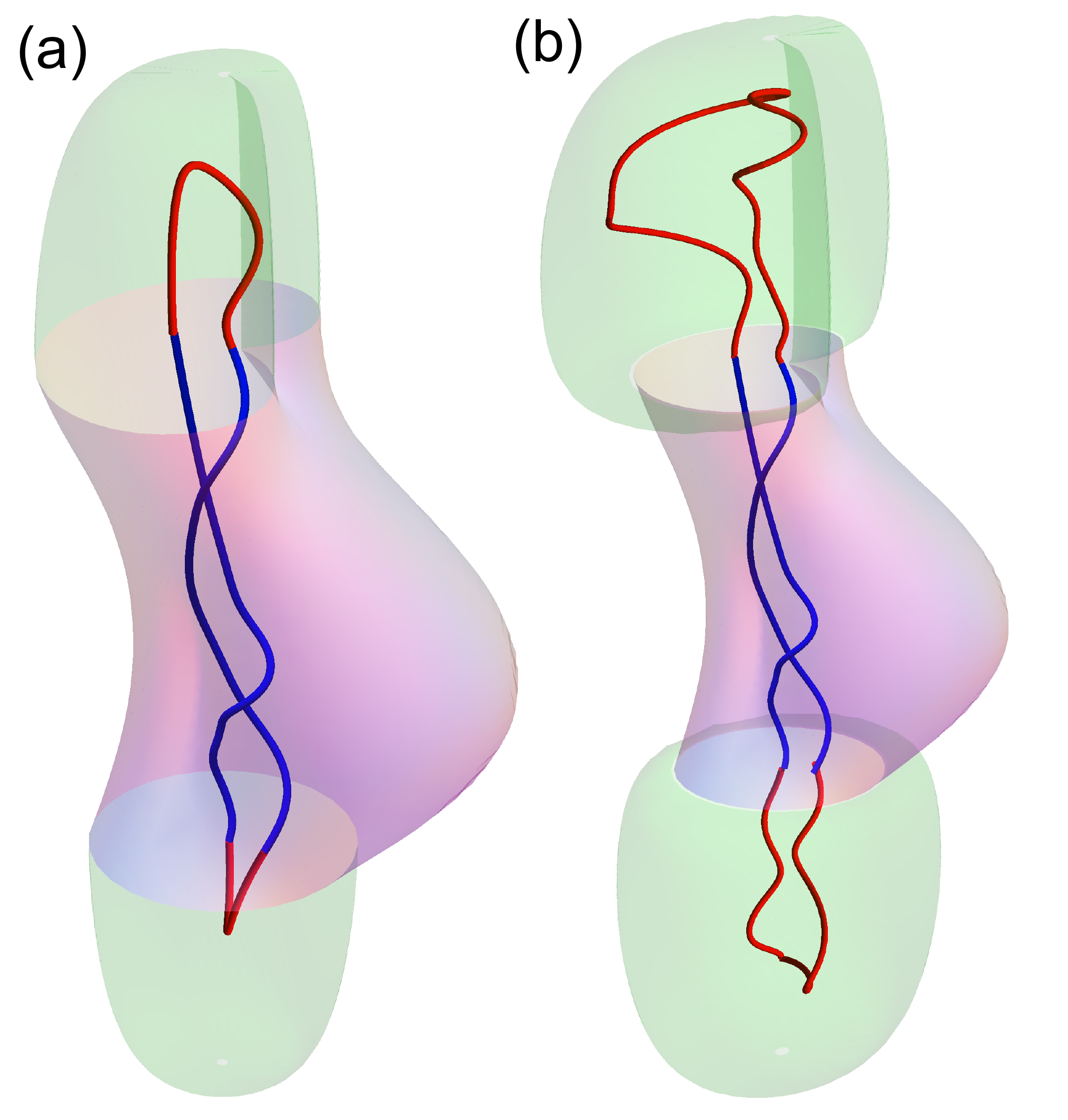}
\caption{The extended volume $V\cup \widetilde{V}$ in the original definition of relative helicity by \citet{Berger1984ay}, for two different choices of extension $\widetilde{V}$. Note that the field lines of $\widetilde{\bf B}$ close the existing field lines.}
\label{fig:hr}
\end{center}
\end{figure}
The original definition of relative helicity for open magnetic fields invokes an imagined extension of $V$ to a larger volume $V\cup \widetilde{V}$ \citep{Berger1984ay}, whose outer boundary is a magnetic surface. Examples of this extended domain are shown in Figure \ref{fig:hr}. We let $\widetilde{\bf B}$ be some magnetic field on $\widetilde{V}$ such that $\widetilde{B}_n$ matches $B_n$ on the boundary of the original volume $V$, and $\widetilde{B}_n$ vanishes on the outer boundary of the combined volume. Let
\begin{equation}
H({\bf B},\widetilde{\bf B}) := \int_V{\bf A}\cdot{\bf B}\,\rmd^3x + \int_{\widetilde{V}}\widetilde{\bf A}\cdot\widetilde{\bf B}\,\rmd^3x, \qquad \textrm{where $\widetilde{\bf B}=\nabla\times\widetilde{\bf A}$.}
\end{equation}
Then the \emph{relative helicity} of ${\bf B}$ with respect to reference field ${\bf B}'$ on $V$ is defined as
\begin{equation}
H_{{\bf B}'}({\bf B}) := H({\bf B},\widetilde{\bf B}) - H({\bf B}',\widetilde{\bf B}),
\end{equation}
for any choice $\widetilde{\bf B}$ of extension field, where ${\bf B}'$ must satisfy the boundary condition $B_n'|_{\partial V}=B_n|_{\partial V}$.

We can show that the relative helicity $H_{{\bf B}'}({\bf B})$ is independent of the choice of extension $\widetilde{\bf B}$, and depends neither on the gauge of ${\bf A}$ nor on that of ${\bf A}'$. To ensure continuity of the vector potential in $H({\bf B},\widetilde{\bf B})$ and $H({\bf B}',\widetilde{\bf B})$, we must use different vector potentials for $\widetilde{\bf B}$ in each case. If we choose ${\bf n}\times\widetilde{\bf A}|_{\partial V}={\bf n}\times{\bf A}|_{\partial V}$ for the first case, then we have ${\bf n}\times\widetilde{\bf A}'={\bf n}\times{\bf A} + {\bf n}\times\nabla\psi$ for the second case. So
\begin{align*}
H_{{\bf B}'}({\bf B}) &= \int_V({\bf A}\cdot{\bf B} - {\bf A}'\cdot{\bf B}')\,\rmd^3x - \int_{\widetilde{V}}\nabla\psi\cdot\widetilde{\bf B}\,\rmd^3x,\\
&= \int_V({\bf A}\cdot{\bf B} - {\bf A}'\cdot{\bf B}')\,\rmd^3x + \oint_{\partial V}\psi B_n\,\rmd^2x.
\end{align*}
Now
\begin{align}
\int_V({\bf A}'\cdot{\bf B} - {\bf A}\cdot{\bf B}')\,\rmd^3x &= \int_V({\bf A}'\cdot\nabla\times{\bf A} - {\bf A}\cdot\nabla\times{\bf A}')\,\rmd^3x,\\
&= \oint_{\partial V}{\bf A}\times{\bf A}'\cdot{\bf n}\,\rmd^2x,\\
&= \oint_{\partial V}{\bf A}\times({\bf A}' - {\bf A})\cdot{\bf n}\,\rmd^2x,\\
&= \oint_{\partial V}{\bf A}\times\nabla\psi\cdot{\bf n}\,\rmd^2x,\\
&= \oint_{\partial V}\Big(\psi\nabla\times{\bf A} - \nabla\times(\psi{\bf A}) \Big)\cdot{\bf n}\,\rmd^2x,\\
&= \oint_{\partial V}\psi B_n\,\rmd^2x. \label{eqn:psibound}
\end{align}
The last line follows from Stokes' Theorem since $\partial V$ is a closed surface. Hence
\begin{equation}
H_{{\bf B}'}({\bf B}) = \int_V ({\bf A} + {\bf A}')\cdot({\bf B} - {\bf B}')\,\rmd^3x.
\label{eqn:finn}
\end{equation}
This is often known as the \citet{Finn1985} formula for relative helicity. Since $\widetilde{\bf B}$ appears nowhere in \eqref{eqn:finn}, we see that $H_{{\bf B}'}({\bf B})$ is independent of the extension $\widetilde{\bf B}$. Gauge invariance readily follows from this formula, for if either ${\bf A}\rightarrow{\bf A}+\nabla\chi$ or ${\bf A}'\rightarrow{\bf A}'+\nabla\chi$, then
\begin{equation}
H_{{\bf B}'}({\bf B}) \rightarrow H_{{\bf B}'}({\bf B}) + \oint_{\partial V}\chi(B_n - B_n')\,\rmd^2x,
\end{equation}
and the last integral vanishes by the boundary condition on ${\bf B}'$.

The main limitation of the relative helicity is that it depends on the choice of reference field ${\bf B}'$, and this complicates its physical interpretation. We return to address this point in Section \ref{sec:hr}.

\section{The Winding Gauge} \label{sec:bs2}

In Section \ref{sec:wind}, we will show that $H$ for an open magnetic field in a domain $D$ such as we consider may be interpreted as an average winding number, analogous to the interpretation of $H$ for a closed magnetic field as an average linking integral. Central to this interpretation will be a specific choice of gauge for ${\bf A}$ that is analogous to the Coulomb gauge of Equation \eqref{eqn:bs3}. The difference in the open case is that the choice of gauge will now affect the value of $H$, not just its integral expression. Nevertheless, this specific gauge - which we call ``winding'' - will turn out to be physically meaningful.

\begin{figure}[t]
\begin{center}
\includegraphics[width=10cm]{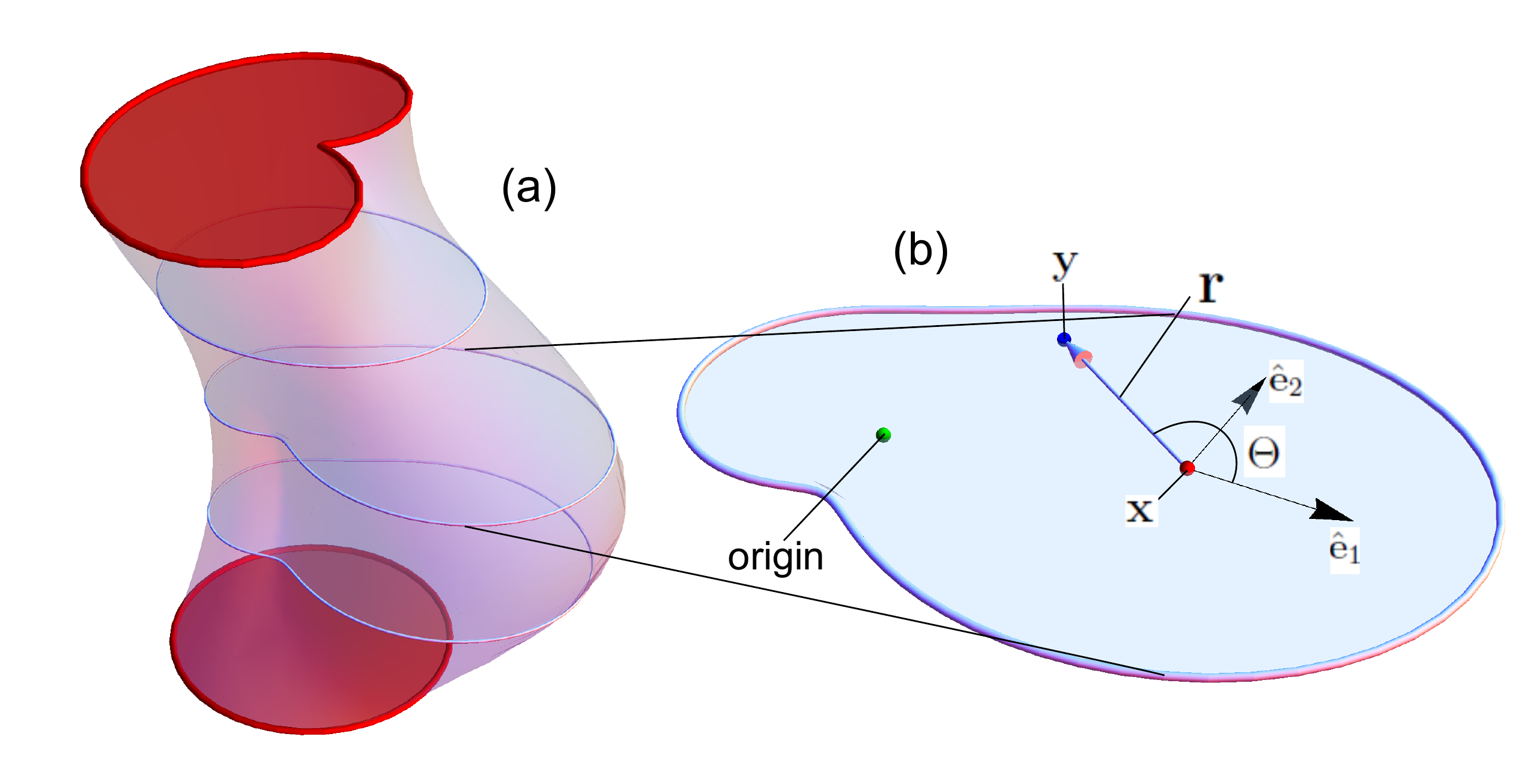}
\caption{The vector ${\bf r}$ used to define the winding gauge and measure the winding of magnetic field lines. The vector potential ${\bf A}^{\rm W}$ at a point on the cross-section $S_z$ (depicted in panel b) is defined by integrating over the cross-section. Also shown is the angle $\Theta$ made by ${\bf r}$ and the ${\bf{\hat{e}}}_1$ axis, and used to define the winding number.}
\label{fig:sz}
\end{center}
\end{figure}

For an open field, we cannot use the Coulomb gauge \eqref{eqn:bs3} since it will generally violate $\nabla\times{\bf A}={\bf B}$ when $B_n|_{\partial V}\neq 0$. However, since $B_n=0$ on the side boundary of our cylinder $V$, it is possible to use a two-dimensional equivalent of the Coulomb gauge whose horizontal divergence $\nabla^\perp\cdot{\bf A}=0$ vanishes (but not its full three-dimensional divergence). This is what we call the \emph{winding gauge}, and may be written
\begin{equation}
{\bf A}^{\rm W}(x_1,x_2,z) = \frac{1}{2\pi}\int_{S_z}\frac{{\bf B}(y_1,y_2,z)\times{\bf r}}{|{\bf r}|^2}\,\rmd^2y, \qquad \textrm{where ${\bf r}=(x_1-y_1,x_2-y_2,0)$}.
\label{eqn:bs2}
\end{equation}
In this gauge, the vector potential at any point is defined as an average over the horizontal surface $S_z$ at that height (Figure \ref{fig:sz}).

Since we have been unable to find it in print, we include here a proof that $\nabla\times{\bf A}^{\rm W}={\bf B}$ for a magnetic field on our domain $V$ with $B_n=0$ on the side boundary. This is similar to the familiar proof for the three-dimensional Coulomb gauge \citep[e.g.,][]{Cantarella2001}, but there are some complications.

Firstly, one may show by direct differentiation that
\begin{equation}
\frac{\bf r}{|{\bf r}|^2} = \nabla^\perp_x\Big(\log|{\bf r}|\Big),
\end{equation}
where $\nabla^\perp_x:=(\partial/\partial x_1, \partial/\partial x_2, 0)$. The subscript $x$ indicates differentiation with respect to $x$, as opposed to $y$. Thus
\begin{align}
{\bf A}^{\rm W}({\bf x}) &= \frac{1}{2\pi}\int_{S_z}{\bf B}({\bf y})\times\nabla_x^\perp\Big(\log\vert {\bf r}\vert\Big)\,\rmd^2y\\
 &= -\frac{1}{2\pi}\nabla_x^\perp\times\left(\int_{S_z}{\bf B}({\bf y})\log\vert {\bf r} \vert\,\rmd^2y\right).\label{bsiden}
\end{align}
To take the $\nabla^\perp_x$ operator outside the integral, we have used the fact that the geometries of the cross-sections $S_z$ vary only as a function of the $z$-coordinate.

By writing out the components explicitly, one can verify for a function ${\bf f}=(f_1,f_2,f_3)$ that
\begin{equation}
\nabla\times\nabla^\perp\times{\bf f} = -(\nabla^\perp)^2{\bf f} + \nabla^\perp(\nabla\cdot{\bf f}),
\label{eqn:curlf}
\end{equation}
where $\nabla$ is the full (3-component) operator, and 
\begin{equation}
\label{functioniden}
(\nabla^\perp)^2{\bf f} =\left(\pder{^2f_1}{x_1^2} + \pder{^2f_1}{x_2^2}, \quad \pder{^2f_2}{x_1^2} + \pder{^2f_2}{x_2^2}, \quad \pder{^2f_3}{x_1^2} + \pder{^2f_3}{x_2^2} \right).
\end{equation}
Applying this to Equation \eqref{bsiden}, we can take the derivatives inside the integral, but for the 3-component operator $\nabla_x$ we must account for the fact that the shape of $S_z$ may vary in $z$. Leibniz' rule adds an extra term depending on ${\bf v}\cdot{\bf n}$, where ${\bf v}({\bf y})={\rmd}{\bf y}/\rmd{z}$ is a ``velocity'' describing how the boundary $S_s$ changes shape in $z$. Since ${\bf B}$ is tangent to $S_s$, we can simply take ${\bf v}={\bf B}/B_z$. Applying \eqref{eqn:curlf} with Leibniz' rule then gives
\begin{align}
\nabla_x\times{\bf A}^{\rm W}({\bf x}) &= \frac{1}{2\pi}\int_{S_z}{\bf B}({\bf y})(\nabla_x^\perp)^2\Big(\log\vert{\bf r}\vert \Big)\,\rmd^2y\nonumber\\
&\quad - \frac{1}{2\pi}\nabla_x^\perp\left(\int_{S_z}\nabla_x\cdot\Big({\bf B}({\bf y})\log\vert{\bf r}\vert \Big)\,\rmd^2y \right)\nonumber\\
& \quad +  \frac{1}{2\pi}\nabla_x^{\perp}\oint_{\partial S_z}B_n({\bf y})\log\vert{\bf r}\vert\,\rmd{l_y}.\label{eqn:curlabs}
\end{align}
For the second term, note that
\begin{align}
\nabla_x\cdot\Big({\bf B}({\bf y})\log\vert{\bf r}\vert \Big) &= {\bf B}({\bf y})\cdot\nabla_x\Big(\log\vert{\bf r}\vert\Big) + (\log\vert{\bf r}\vert)\nabla_x\cdot{\bf B}({\bf y}),\\
&= -{\bf B}({\bf y})\cdot\nabla_y\Big(\log\vert{\bf r}\vert\Big) + (\log\vert{\bf r}\vert)\pder{B_z({\bf y})}{z},\\
&= -{\bf B}({\bf y})\cdot\nabla^\perp_y\Big(\log\vert{\bf r}\vert\Big) - (\log\vert{\bf r}\vert)\nabla_y^\perp\cdot{\bf B}(y),\\
&= -\nabla_y^\perp\cdot\Big({\bf B}({\bf y})\log\vert{\bf r}\vert  \Big).
\end{align}
The resulting term becomes a boundary integral and the sum of the second and third terms of (\ref{eqn:curlabs}) cancel, leaving just the first term. It may be shown that
\begin{equation}
(\nabla^\perp_x)^2\Big(\log\vert{\bf r}\vert\Big) = 2\pi\delta({\bf r}),
\end{equation}
where $\delta({\bf r})$ is the two-dimensional Dirac $\delta$-function and $(\nabla^\perp_x)^2$ is the two-dimensional Laplacian. (This is a direct analogue of a widely-used result in three dimensions.) It follows that $\nabla_x\times{\bf A}^{\rm W}({\bf x})={\bf B}({\bf x})$ so that \eqref{eqn:bs2} is a valid vector potential.

\section{Winding Number Interpretation of Helicity} \label{sec:wind}

For a pair of field lines ${\bf x}(s)$, ${\bf y}(t)$ that are not closed within $V$, the Gauss linking number is not a topological invariant. Instead, we shall define the \emph{winding number} ${\cal L}({\bf x},{\bf y})$ between two field lines in $z$.

\begin{figure}[t]
\begin{center}
\includegraphics[width=\textwidth]{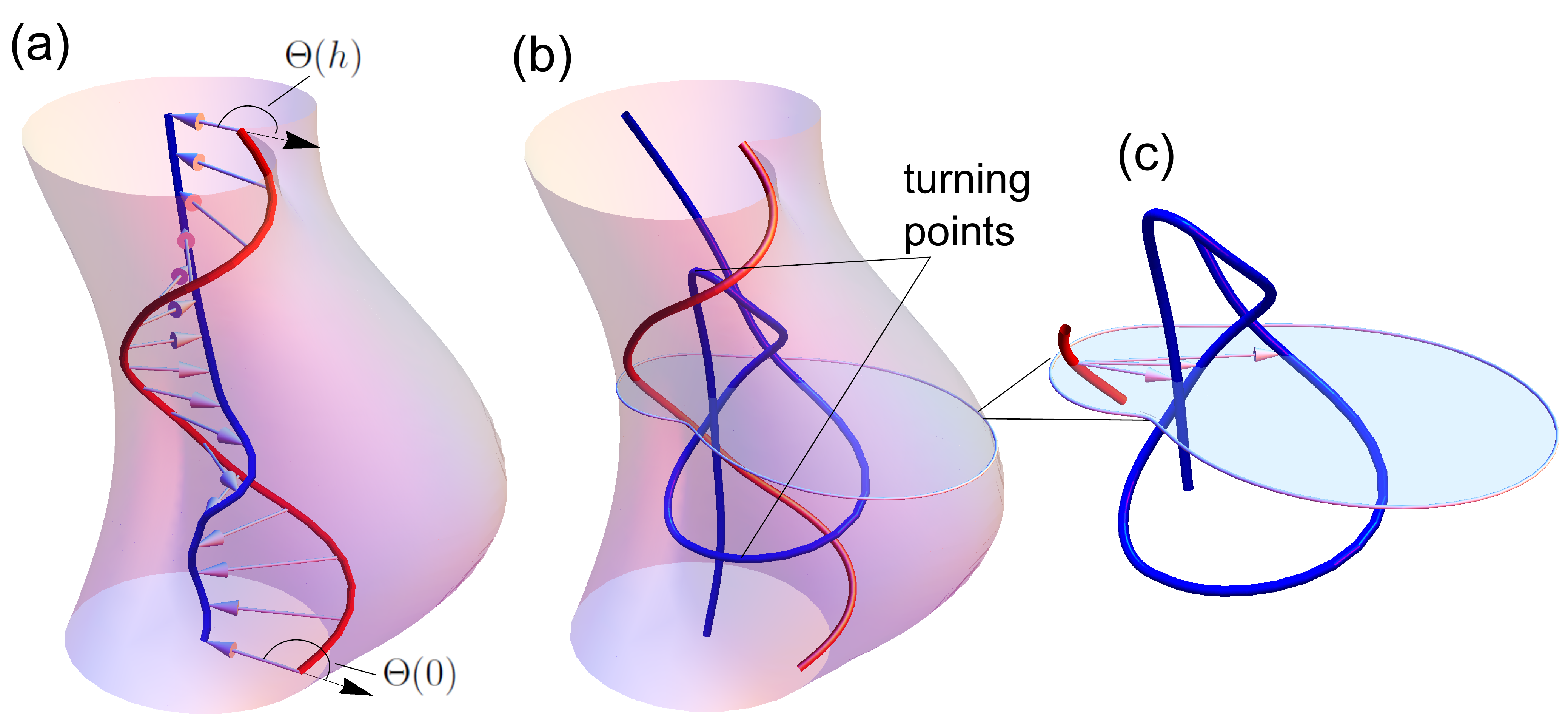}
\caption{Geometrical interpretation of the winding number. Panel (a) depicts a pair of curves whose $z$-components are monotonic in $z$. The vectors ${\bf r}(z)$ joining the two curves are shown, along with the angle $\Theta$ between ${\bf r}$ and the $x$-axis at either end. Panel (b) depicts a pair curves for which we need to define multiple angles $\Theta_{ij}$, owing to the fact that one curve is not monotonic in $z$. Panel (c) depicts an example cross-section from Panel (b), showing three vectors ${\bf r}_{1j},j = 1,2,3$ defining angles on a particular plane $S_z$.}
\label{fig:wind1}
\end{center}
\end{figure}

First, consider the case where the $z$-coordinates of both field lines are monotonically increasing, as in Figure \ref{fig:wind1}(a). In this case, we can parametrize both field lines by their $z$-coordinate. We define ${\cal L}$ to be the net rotation of the vector ${\bf r}$ between ${\bf x}$ and ${\bf y}$ as $z$ increases from $0$ to $h$, so
\begin{equation}
{\cal L}({\bf x},{\bf y}) := \frac{1}{2\pi}\int_0^h\frac{\rmd}{\rmd z}\Theta\big({\bf x}(z),{\bf y}(z)\big)\,\rmd z, \qquad \textrm{where} \quad \Theta({\bf x},{\bf y}) = \arctan\left(\frac{x_2-y_2}{x_1-y_1} \right).
\label{eqn:cflat}
\end{equation}
As $\Theta$ is a multi-valued function, the boundary values on $S_0$ and $S_h$ define ${\cal L}$ only up to an integer, \textit{i.e.},
\begin{equation}
{\cal L}({\bf x},{\bf y}) = \frac{1}{2\pi}\Big(\Theta\big({\bf x}(h),{\bf y}(h)\big) - \Theta\big({\bf x}(0),{\bf y}(0)\big)\Big) + N, 
\end{equation}
where $N$ is the integer number of full windings of the joining vector ${\bf r} = {\bf y}- {\bf x}$.
If the end angles remain fixed, and the field lines remain monotonic in $z$, the winding number is invariant to deformations of the field lines which forbid their crossing \citep{Berger1993b,Berger2006}.

In general, the field lines ${\bf x}$ and ${\bf y}$ may have respectively $n$ and $m$ distinct points in $z$ where they turn back on themselves, that is $\rmd x_z/\rmd z=0$ or $\rmd y_z/\rmd z=0$. We split ${\bf x}$ into $n+1$ sections at these turning points and similarly split ${\bf y}$ into $m+1$ sections. For example the blue field line in Figure \ref{fig:wind1}(b) has two such turning points, so is split into three sections. Sections ${\bf x}_i$ and ${\bf y}_j$ share a mutual $z$-range $[z_{ij}^{min},z_{ij}^{max}]$ (which could be an empty set), and for each of these ranges we can define an angle $\Theta({\bf x}_i,{\bf y}_j)$ for each vector ${\bf r}_{ij}$. For example, there are three such vectors between the blue and red curves in Figure \ref{fig:wind1}(c), because the blue curve has three sections passing through this plane. \cite{Berger2006} defined the sum
\begin{equation}
\label{windingnumber}
{\cal L}({\bf x},{\bf y}) := \sum_{i=1}^{n+1}\sum_{j=1}^{m+1}\frac{\sigma({\bf x}_i)\sigma({\bf y}_j)}{2\pi}\int_{z_{ij}^{min}}^{z_{ij}^{max}}\frac{\rmd \Theta\big({\bf x}_i(z),{\bf y}_j(z)\big)}{\rmd z}\rmd{z}.
\end{equation}  
where $\sigma({\bf x}_i)$ is an indicator function marking whether the curve section ${\bf x}_i$ moves up or down in $z$; for example
\begin{equation}
\sigma({\bf x}_i) = \left\{
\begin{array}{cc}
1 & \mbox{ if } {\rmd x_z}/{\rmd z} >0,\\
-1 & \mbox{ if } {\rmd x_z}/{\rmd z} <0.
\end{array}
\right.
\end{equation}
It was shown that under this extended definition that ${\cal L}({\bf x},{\bf y})$ remains invariant to all deformations which vanish at the bounding planes $S_0$, $S_h$ and forbid self-crossings. Importantly, this remains true whether one or both field lines are anchored only at one plane, or are closed in $V$ \citep{Berger2006}. For example, it could be applied to any pair of curves in Figure \ref{fig:defs}(c). It was further shown that for closed curves ${\cal L}$ is has the same integer value as the Gauss linking integral \eqref{gl}. We re-iterate that the winding number is not equal to the Gauss linking integral for open curves. Indeed, the linking number is not an invariant in such cases, so the winding number framework for topological classification is applicable to a much larger set of admissible magnetic fields.

\subsection{Winding Gauge}
Our goal in this Section is to express $H$ as an average pairwise winding between the field lines that make up the open magnetic field. Firstly, we will show this for the winding gauge of Section \ref{sec:bs2}, then we will consider what happens in a general gauge.

If ${\bf A}$ is written in the winding gauge \eqref{eqn:bs2}, we will show that the corresponding helicity $H$, which we shall denote $\Hw$, is related to the winding numbers by
\begin{equation}
\Hw({\bf B}) := \int_V{\bf A}^{\rm W}\cdot{\bf B}\,d^3x = \frac{1}{2\pi}\int_{0}^{h}\int_{S_z\times S_z}\frac{\rmd}{\rmd z}\Theta\big({\bf x},{\bf y} \big)B_z({\bf x})B_z({\bf y})\,\rmd^2x\,\rmd^2y\,\rmd{z}.
\label{eqn:hbs}
\end{equation}
So the helicity is the average pairwise winding between all local portions of field lines. To relate $\Hw$ to the winding of entire field lines, consider the flux-weighted winding number
\begin{equation}
{\cal L}_{\bf B}({\bf x},{\bf y}) := \sum_{i=1}^{n+1}\sum_{j=1}^{m+1}\frac{1}{2\pi}\int_{z_{ij}^{min}}^{z_{ij}^{max}}\frac{\rmd \Theta\big({\bf x}_i(z),{\bf y}_j(z)\big)}{\rmd z}B_z({\bf x}_i)B_z({\bf y}_j)\rmd{z}.
\end{equation} 
Like ${\cal L}$, this is invariant under deformations that vanish on $S_0$, $S_h$ and and forbid self crossings. (The $B_z$ functions have the same sign as the $\sigma$ functions in \eqref{windingnumber}.) Then $\Hw$ may be formally written as the average flux-weighted winding number over all pairs of field lines ${\bf x}(s)$, ${\bf y}(s)$, or
\begin{equation}
\Hw({\bf B}) = \int {\cal L}_{\bf B}({\bf x},{\bf y})\,\rmd{\bf x}\,\rmd{\bf y}.
\end{equation}
In the particular case where the $z$-coordinates of all field lines are monotonically increasing, $\Hw$ may be written as an integral of the original winding number ${\cal L}({\bf x},{\bf y})$ over $S_0$,
\begin{equation}
\Hw({\bf B}) = \int_{S_0}{\cal L}({\bf x},{\bf y})B_z({\bf x})B_z({\bf y})\,\rmd^2x\,\rmd^2y.
\end{equation}
In this case, one can also write
\begin{equation}
\Hw({\bf B}) = \int_{S_0}{\cal A}^{\rm W}({\bf x})B_z({\bf x})\,\rmd^2x,
\end{equation}
where the \emph{flux function}
\begin{equation}
{\cal A}^{\rm W}({\bf x}) := \int_{{\bf x}(z)}\frac{{\bf A}^{\rm W}\cdot{\bf B}}{|B_z|}\,\rmd z
\label{eqn:fluxfun}
\end{equation}
is obtained by integrating the vector potential over a particular field line ${\bf x}(z)$.  Although we do not dwell on it in this Paper, notice that the flux function ${\cal A}^{\rm W}$ is also a physically meaningful ideal invariant. In fact for fields with $B_z>0$ everywhere it has been shown to be more powerful than the helicity $\Hw$ and able to distinguish between 
topologically different magnetic fields with the same $\Hw$ \citep{yeates2013}.

To prove Equation \eqref{eqn:hbs}, consider the helicity density
\begin{align}
{\bf A}^{\rm W}\cdot {\bf B} &= {\bf B}({\bf x})\cdot \left(\frac{1}{2\pi}\int_{S_z}\frac{{\bf B}({\bf y})\times{\bf r}}{|{\bf r}|^2}\,\rmd ^2y\right),\\
 &=\frac{1}{2\pi}\left({\bf B}^\perp({\bf x})\cdot\int_{S_z}B_z({\bf y})\frac{\hat{\bf z}\times{\bf r}}{|{\bf r}|^2}\,\rmd^2y - B_z({\bf x})\int_{S_z}{\bf B}^\perp({\bf y})\cdot\frac{\hat{\bf z}\times{\bf r}}{|{\bf r}|^2}\,\rmd^2y\right),\\
&= \frac{1}{2\pi}\int_{S_z}\left(\frac{{\bf B}^\perp({\bf x})}{B_z({\bf x})} - \frac{{\bf B}^\perp({\bf y})}{ B_z({\bf y})}\right)\cdot\frac{\hat{\bf z}\times{\bf r}}{|{\bf r}|^2}B_z({\bf y})B_z({\bf x})\,\rmd^2y.
\end{align}
Now, differentiating $\Theta({\bf x},{\bf y})$ with respect to $z$ yields
\begin{align}
\frac{\rmd}{\rmd z}\Theta\big({\bf x}(z),{\bf y}(z)\big) &= \frac{r_1^2}{\vert{\bf r}\vert^2}\frac{\rmd}{\rmd z}\left(\frac{r_2}{r_1}\right),\\
 &= \frac{1}{\vert{\bf r}\vert^2}\left(r_1\frac{\rmd r_2}{\rmd z} - r_2\frac{\rmd r_1}{\rmd z} \right),\\
 &= \frac{1}{\vert{\bf r}\vert^2}\left\{r_1\left(\frac{B_2({\bf x})}{B_z({\bf x})} - \frac{B_2({\bf y})}{B_z({\bf y})}\right) - r_2\left(\frac{B_1({\bf x})}{B_z({\bf x})} - \frac{B_1({\bf y})}{B_z({\bf y})}\right) \right\}.
\end{align}
Here we have used that ${\bf x}(z)$, ${\bf y}(z)$ are segments of magnetic field lines. Since $\hat{\bf z}\times{\bf r}=(-r_2,r_1,0)$, we arrive at
\begin{equation}
{\bf A}^{\rm W}\cdot {\bf B}  = \frac{1}{2\pi}\int_{S_z}\frac{\rmd}{\rmd z}\Theta\big({\bf x},{\bf y}\big)B_z({\bf y})B_z({\bf x})\,\rmd^2y,
\end{equation}
and integrating over $V$ gives Equation \ref{eqn:hbs}.

It is important to notice that Equation \eqref{eqn:hbs} defines $\Hw({\bf B})$ uniquely using only ${\bf B}$ itself, without reference to ${\bf A}$. This reinforces the fact that $\Hw$ is a physically meaningful quantity, despite the absence of an explicit reference field. The interpretation of $\Hw$ as an average pairwise winding number is analogous to the interpretation of $H$ for a closed field as a flux-weighted average of the linking number between all pairs of field lines. The fact that the winding number ${\cal L}({\bf x},{\bf y})$ of a closed curve is equal to the linking number \eqref{gl} mirrors the fact that, for a closed field, $\Hw$ matches the value of $H$ obtained with the Coulomb gauge \eqref{eqn:bs3}, owing to the gauge independence of $H$ for closed fields.

One last note in this section is the fact that the winding number expression \eqref{eqn:hbs} for the topology of open fields was obtained by \cite{Berger1986} for a field in the half space (decaying at a sufficient rate towards infinity). In that case it was attributed to the \emph{relative helicity} $H_{{\bf B}^{\rm p}}({\bf B})$ with a potential reference field ${\bf B}^{\rm p}$. As well shall see in Section \ref{sec:aper}, the identification of $H_{{\bf B}^{\rm p}}({\bf B})$ with average winding is not always true for the more general set of fields we consider here, whilst we have seen that $\Hw$ always has this interpretation.

\subsection{Other Gauges} \label{sec:gen}

What is the physical meaning of $H$ in a gauge other than the winding gauge? In such a gauge we have ${\bf A}'={\bf A}^{\rm W} + \nabla\chi$ for some scalar function $\chi$. It turns out that the choice of $\chi$ corresponds to a particular choice of \emph{frame field} for defining the angle $\Theta$.

In \eqref{eqn:cflat}, we defined $\Theta$ as $\arctan(r_2/r_1)$, where $r_1$, $r_2$ are the components of ${\bf r}$ with respect to a particular orthonormal Cartesian frame $\{{\bf \hat{e}}_1,{\bf \hat{e}}_2\}$. But suppose we choose a different orthonormal frame $\{{\bf \hat{e}}_1',{\bf \hat{e}}_2'\}$, rotated through angle $\theta$ with respect to $\{{\bf \hat{e}}_1,{\bf \hat{e}}_2\}$. Then the components of ${\bf r}$ with respect to the new frame are
\begin{equation}
r_1'=r_1\cos\theta - r_2\sin\theta, \qquad r_2'=r_1\sin\theta + r_2\cos\theta.
\end{equation}
Defining the angle with respect to this new frame gives a different result
\begin{equation}
\Theta' := \arctan\left(\frac{r_2'}{r_1'}\right) = \arctan\left(\frac{r_2/r_1 + \tan\theta}{1 - (r_2/r_1)\tan\theta}\right) = \Theta + \theta.
\end{equation}
On a particular cross-section $S_z$ the new winding rate relates to the old winding rate through 
\begin{equation}
\label{eqn:cprden}
\frac{\rmd}{\rmd z}\Theta'({\bf x},{\bf y}) = \frac{\rmd}{\rmd z}\Theta({\bf x},{\bf y}) + \frac{\rmd}{\rmd z}\theta({\bf x}),  
\end{equation}
and the new winding number of the curves ${\bf x}$ and ${\bf y}$ is
\begin{equation}
{\cal L}_{\bf B}'({\bf x},{\bf y}) = {\cal L}_{\bf B}({\bf x},{\bf y}) + \sum_{i=1}^{n+1}\sum_{j=1}^{m+1}\frac{1}{2\pi}\int_{z_i^{min}}^{z_i^{max}}\frac{\rmd}{\rmd z}\theta({\bf x}_i)B_z({\bf x}_i)B_z({\bf y}_j)\,\rmd z.
\label{eqn:cpr}
\end{equation}
This reduces to ${\cal L}_{\bf B}({\bf x},{\bf y})$ if $\theta$ is constant ({\it i.e.}, if we always measure $\Theta$ with respect to the same frame). But if there is a net change in $\theta$ along the field line ${\bf x}(z)$, then the new winding number ${\cal L}_{\bf B}'({\bf x},{\bf y})$ differs from the old ${\cal L}_{\bf B}({\bf x},{\bf y})$.  It should be pointed out that the new winding measure remains invariant to ideal motions, but now part of its value is due to a non-physical quantity: the rotation of the frame field. An example will be shown in Section \ref{sec:ex}.

To see that our frame field corresponds to a change of gauge, we can substitute \eqref{eqn:cprden} into \eqref{eqn:hbs} to find that
\begin{align}
&\frac{1}{2\pi}\int_{0}^{h}\int_{S_z\times S_z}\frac{\rmd}{\rmd z}\Theta'({\bf x},{\bf y}) B_z({\bf x}) B_z({\bf y})\,\rmd^2x\,\rmd^2y\, \rmd{z}\nonumber\\
\qquad\qquad &= \Hw({\bf B})
+ \frac{\Phi_0}{2\pi}\int_0^h\int_{S_z}\frac{\rmd}{\rmd z}\theta({\bf x})B_z({\bf x})\,\rmd^2x\,\rmd z.
\end{align}
Using Equation \eqref{eqn:hgauge}, and if the gauge function $\chi$ satisfies
\begin{equation}
\int_{S_h}\chi B_z \,d^2x - \int_{S_0}\chi B_z \,d^2x = \frac{\Phi_0}{2\pi}\int_0^h\int_{S_z}\frac{\rmd\theta}{\rmd z}B_z\,\rmd^2x\,\rmd z,
\end{equation}
if follows that
\begin{equation}
H'({\bf B}) = \frac{1}{2\pi}\int_{0}^{h}\int_{S_z\times S_z}\frac{\rmd}{\rmd z}\Theta'({\bf x},{\bf y}) B_z({\bf x}) B_z({\bf y})\,\rmd^2x\,\rmd^2y\, \rmd{z}.
\label{eqn:hbsp}
\end{equation}
For example, given the frame field $\theta({\bf x})$, one could take the gauge function
\begin{equation}
\chi(x_1,x_2,z) = \frac{z}{2\pi h}\int_0^h\int_{S_z}\frac{\rmd}{\rmd z}\theta({\bf y})B_z({\bf y})\,\rmd^2y\,\rmd z,
\end{equation}
but there are many possible gauges that will give the same helicity as a particular frame field.

In summary, Equation \eqref{eqn:hbsp} shows that the helicity in an arbitrary gauge is still the average pairwise winding number, but now the winding number is measured with respect to a frame field $\theta({\bf x})$ that varies in space. In most situations, it seems more physically meaningful to measure winding with respect to a fixed frame, in which case the winding gauge is most appropriate. A different choice of gauge corresponds to measuring winding with respect to a varying frame, whereby even a straight magnetic field may appear tangled. This is highlighted in the following example.

\subsection{Example} \label{sec:ex}

\begin{figure}[bt]
\begin{center}
\includegraphics[width=0.7\textwidth]{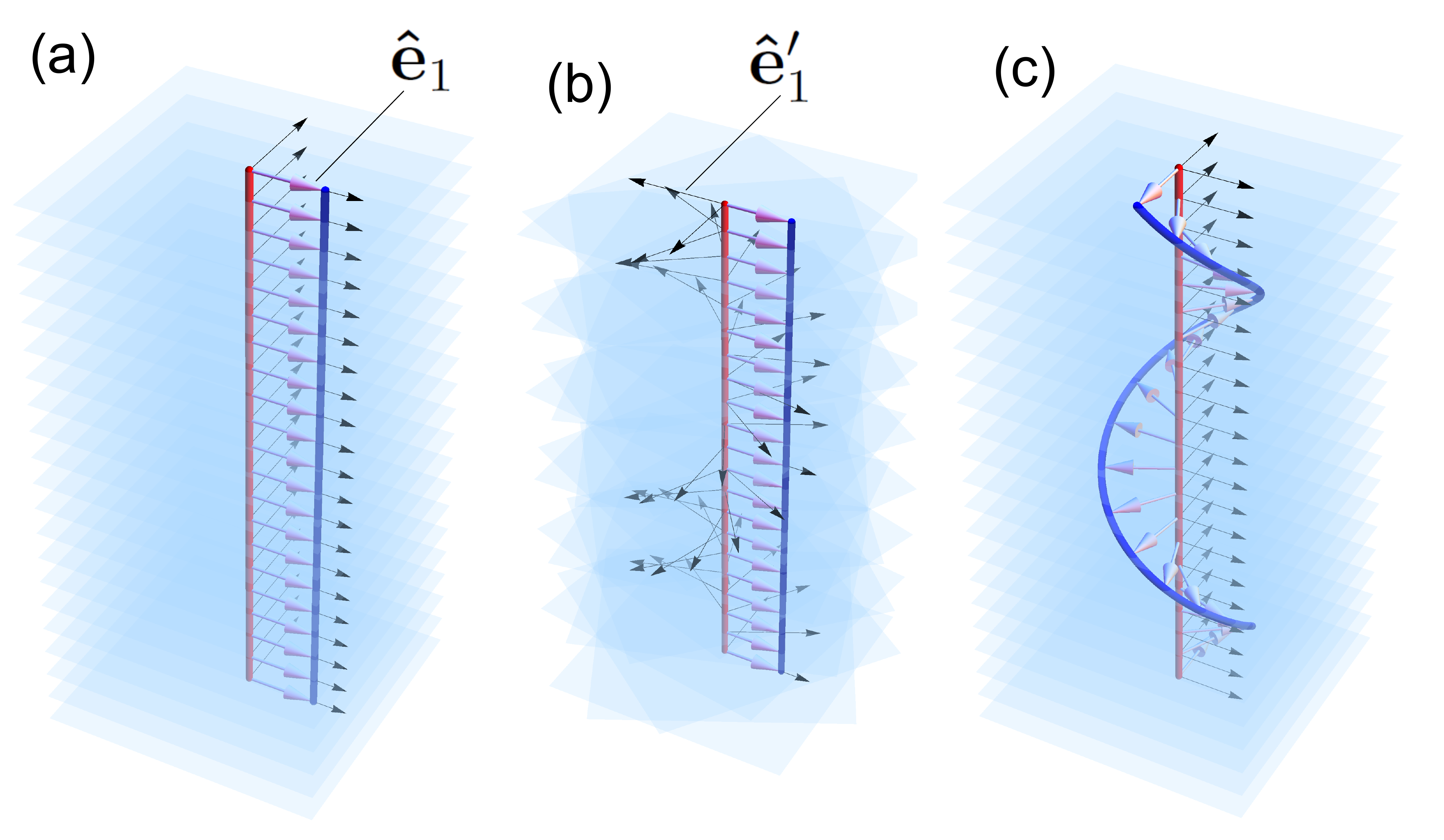}
\caption{\label{gengauge} How a straight field can appear twisted when measured with respect to a rotating frame. Panel (a) depicts a pair of straight field lines (from a field ${\bf B}=\hat{\bf z}$) and their joining vectors ${\bf r}(z)$, which have no winding. Panel (b) shows the same figure with a varying basis indicated by the black rotating arrows. The angle between ${\bf r}$ and $\hat{\bf e}'_1$ in this basis would rotate with $z$. Panel (c) shows two field lines of a helical field  ${\bf B}=\hat{\bf z} + (3\pi/R_0)r{\bf e}_\phi$, one of which is straight (at $r=0$). The pairwise winding of these two field lines in the original basis is the same as that of the straight field in the rotated basis.}
\label{fig:frame}
\end{center}
\end{figure}

To illustrate the idea of Section \ref{sec:gen}, consider the uniform vertical field ${\bf B}=\hat{\bf z}$, in a circular cylinder of radius $R_0$ and height $h$. One may show by direct calculation from \eqref{eqn:bs2} that the winding-gauge vector potential of this field is ${\bf A}^{\rm W}(r,\phi,z)=(r/2)\hat{\bf e}_\phi$ in standard cylindrical coordinates, and hence that $\Hw({\bf B})=0$. This is consistent with the fact that all field lines of ${\bf B}$ are vertical and untwisted, so that all pairwise winding numbers ${\cal L}({\bf x},{\bf y})$ vanish when measured with respect to a fixed frame (Figure \ref{gengauge}a).

However, suppose that we measure the winding numbers with respect to a frame field $\theta(z)$ that varies in $z$ but (for simplicity) not in $r$, $\phi$. Then each pair of field lines have the same non-zero winding number 
\begin{equation}
{\cal L}'({\bf x},{\bf y})= {\cal L}_0 := \frac{1}{2\pi}\int_0^h\frac{\rmd}{\rmd z}\theta(z)\,\rmd z,
\end{equation}
proportional to the net rotation of the frame. According to Section \ref{sec:gen}, this corresponds to measuring $H$ with the vector potential ${\bf A}^{\rm W}+\nabla\chi$, where
\begin{equation}
\chi(z) = \frac{z}{2\pi h}\int_0^h\int_{S_z}\frac{\rmd}{\rmd z}\theta(z)B_z\,\rmd^2x\,\rmd z = \frac{\pi R_0^2z}{h}{\cal L}_0.
\end{equation}
In this new gauge, $H({\bf B})$ will take the non-zero value $H'({\bf B})=(\pi R_0^2)^2{\cal L}_0$. For example, in Figure \ref{gengauge}(b) we have chosen to rotate the frame by $\theta(z) = 3\pi z$ (hence if $h=1$,  ${\cal L}_0 = 3/2$). In fact, this is exactly the helicity one would get using the winding gauge but for a uniformly-twisted magnetic field ${\bf B}=\hat{\bf z} + (2\pi {\cal L}_0/R_0)r\hat{\bf e}_\phi$ (cf. the calculation in Appendix \ref{app:ref}), whose field lines are helices, as depicted in Figure \ref{gengauge}(c).

In other words, by defining winding with respect to a twisted frame field $\theta$, our straight magnetic field appears twisted. If we choose an arbitrary gauge to define $H$, we are effectively changing our definition of ``untwisted''. The winding gauge is the natural choice because then $\Hw$ is measuring the net winding with respect to a straight field. We feel that this is an important issue to highlight as it provides a geometrical insight into the meaning of the choice of gauge, and consequently a clear reason for showing preference to a particular gauge. If one were asked to measure the winding of a pair of field lines, it would be unnatural to measure the angle made by the two curves $\Theta$ with respect to anything but a fixed frame; yet, as we shall see in the following section, defining the relative helicity with a potential reference field is in many cases equivalent to choosing a varying frame whose net rotation is non-zero.

\section{Comparison with Relative Helicity} \label{sec:hr}

We have shown that the magnetic helicity $H$ of an open magnetic field may be physically interpreted as an average winding. Changing the gauge reflects a change in how the winding numbers are measured, but there is a unique gauge that measures winding with respect to a fixed frame: the winding gauge introduced in Section \ref{sec:bs2}. In this Section, we compare the corresponding helicity $\Hw$ with the relative helicity $H_{{\bf B}'}$ that is typically used in solar physics. This comparison helps us to address the question of whether the potential field ${\bf B}^{\rm p}$ is ``untwisted'' (in the sense of $\Hw({\bf B}^{\rm p})=0$), and leads to a proof of the ``equivalence'' of standard and relative helicity.

\subsection{General relation} 

Let $\Hw({\bf B})$ denote the winding helicity of ${\bf B}$ as defined in Equation \eqref{eqn:hbs}, and let $H_{{\bf B}'}({\bf B})$ be the relative helicity with some reference field ${\bf B}'$. Then we claim that
\begin{equation}
H_{{\bf B}'}({\bf B}) = H^{{\rm W}}({\bf B}) - H^{{\rm W}}({\bf B}').
\label{eqn:rbs}
\end{equation}
In other words, the cross term in Equation \eqref{eqn:finn} vanishes if both ${\bf A}$ and ${\bf A}'$ are written in winding gauge. Note that the winding gauge is not the only gauge for which the cross-term in \eqref{eqn:finn} vanishes; for example, this property is shared by the cylindrical helicity of \citet{Low2011h} and the gauge choice of \citet{Valori2012b}.
To prove \eqref{eqn:rbs} for the winding gauge, note that
\begin{align}
\int_V{\bf A}'\cdot{\bf B}\,d^3x &= \frac{1}{2\pi}\int_0^h\int_{S_z\times S_z}{\bf B}({\bf x})\cdot\frac{{\bf B}'({\bf y})\times{\bf r}}{|{\bf r}|^2}\,\rmd^2y\,\rmd^2x\,\rmd z,\\
&= -\frac{1}{2\pi}\int_0^h\int_{S_z\times S_z}{\bf B}'({\bf y})\cdot\frac{{\bf B}({\bf x})\times{\bf r}}{|{\bf r}|^2}\,\rmd^2y\,\rmd^2x\,\rmd z,\\
&= \frac{1}{2\pi}\int_0^h\int_{S_z\times S_z}{\bf B}'({\bf x})\cdot\frac{{\bf B}({\bf y})\times{\bf r}}{|{\bf r}|^2}\,\rmd^2y\,\rmd^2x\,\rmd z,\\
&= \int_V{\bf A}\cdot{\bf B}'\,d^3x.
\end{align}
Of course, the choice of winding gauge for either ${\bf A}$ or ${\bf A}'$ does not change the value of the relative helicity: only the choice of reference field ${\bf B}'$ does this. It is clear from \eqref{eqn:rbs} that choosing a reference field with vanishing $\Hw({\bf B}')$ will make the relative helicity equal to $\Hw({\bf B})$, so that it inherits the same physical meaning.

\subsection{Untwisted Reference Fields} \label{sec:untwist}

A magnetic field with $\Hw({\bf B})=0$ might be described as ``untwisted'' in a well-defined physical sense. If we use such a field as the reference field in the relative helicity, then Equation \eqref{eqn:rbs} shows that the relative helicity reduces to $\Hw$. It is therefore interesting to find that the most commonly used reference field - the potential field in $V$ - does not always satisfy $\Hw({\bf B})=0$. In this section we assume (for concreteness) that $V$ is a circular cylinder $\{0\leq r\leq R_0, \, 0\leq\phi < 2\pi, \, -L\leq z\leq L \}$. For consistency, we continue to denote the lower and upper boundaries by $S_0$ and $S_h$, and the side boundary by $S_s$.

\subsubsection{Periodic Boundary Conditions} \label{sec:periodic}

For a \emph{periodic} magnetic field, i.e. when $B_z(r,\phi,L)=B_z(r,\phi,-L)$, a simple choice of reference field with $H^{{\rm W}}({\bf B}')=0$ is the vertical field ${\bf B}^{\rm v}(r,\phi,z) = B_z(r,\phi,-L)\hat{\bf z}$, which is readily seen to have $\Hw({\bf B}^{\rm v})=0$. We can use ${\bf B}^{\rm v}$ to prove that the potential field ${\bf B}^{\rm p}$ also has $\Hw({\bf B}^{\rm p})=0$ for a periodic field.

Our strategy is to prove that $H_{{\bf B}^{\rm v}}({\bf B}^{\rm p})=0$. Since we know that $\Hw({\bf B}^{\rm v})=0$, we can then use Equation \eqref{eqn:rbs} to conclude that $\Hw({\bf B}^{\rm p})=0$. To calculate $H_{{\bf B}^{\rm v}}({\bf B}^{\rm p})$, let
\begin{equation}
{\bf B}^{\rm p} = \nabla\left(\pder{\psi}{z} + B_0z\right), \qquad \textrm{where $\nabla^2\psi=0$.}
\label{eqn:pot}
\end{equation} 
This representation is general and lets us write the vector potential (in cylindrical coordinates) as
\begin{equation}
{\bf A}^{\rm p}=\nabla\times(\psi\hat{\bf z}) + \frac{rB_0}{2}{\bf e}_\phi.
\end{equation}
(A similar representation in spherical coordinates was used by \citealp{vanBallegooijen2000}.) For the reference field, we choose the gauge ${\bf A}^{\rm v}(r,\phi,z)=A^{\rm p}_\phi(r,\phi,-L){\bf e}_\phi$. In these gauges, we have
\begin{equation}
H_{{\bf B}^{\rm v}}({\bf B}^{\rm p}) = \int_V{\bf A}^{\rm p}\cdot{\bf B}^{\rm p}\,\rmd^3x + \oint_{\partial V}{\bf A}^{\rm p}\times{\bf A}\cdot{\bf n}\,\rmd^3x.
\label{eqn:hvp}
\end{equation}
The first term gives
\begin{align}
\int_V{\bf A}^{\rm p}\cdot{\bf B}^{\rm p}\,d^3x &= \int_V\left(\nabla\times(\psi\hat{\bf z}) + \frac{rB_0}{2}{\bf e}_\phi \right)\cdot\nabla\left(\pder{\psi}{z} + B_0z \right)\,\rmd^3x,\\
&= \oint_{\partial V}\left(\pder{\psi}{z} + B_0z \right){\bf n}\cdot(\nabla\psi)\times\hat{\bf z}\,\rmd^2x,\\
&= \int_{S_s}\left(\pder{\psi}{z} + B_0z \right)\pder{\psi}{\phi}\,\rmd\phi\,\rmd z,\\
&= \int_{S_s}\pder{\psi}{z}\pder{\psi}{\phi}\,\rmd\phi\,\rmd z + B_0\int_{-L}^L z\left(\int_0^{2\pi}\pder{\psi}{\phi}\,d\phi\right)\,\rmd z.
\end{align}
The last integral vanishes by periodicity in $\phi$. The surface integral in \eqref{eqn:hvp} vanishes on $S$ thanks to our choices of gauge, leaving
\begin{align}
\oint_{\partial V}{\bf A}^{\rm p}\times{\bf A}\cdot{\bf n}\,\rmd^3x &= \int_{S_h-S_0}A^{\rm p}_rA^{\rm v}_\phi\,\rmd^2x,\\
&= \int_{S_h-S_0}A^{\rm p}_rA^{\rm p}_\phi\,\rmd^2x,\\
&= \int_{S_h-S_0}\frac{1}{r}\pder{\psi}{\phi}\left(-\pder{\psi}{r} + \frac{rB_0}{2}\right)r\,\rmd\phi\,\rmd r,\\
&= - \int_{S_h-S_0}\pder{\psi}{\phi}\pder{\psi}{r}\,\rmd\phi\,\rmd r + \frac{B_0}{2}\int_0^{R_0}r\left(\int_0^{2\pi}\pder{\psi}{\phi}\,\rmd\phi \right)\,\rmd r.
\end{align}
Again the last term vanishes by periodicity in $\phi$. Overall, we are left with
\begin{equation}
H_{{\bf B}^v}({\bf B}^{\rm p}) =\int_{S_s}\pder{\psi}{z}\pder{\psi}{\phi}\,\rmd\phi\,\rmd z - \int_{S_h-S_0}\pder{\psi}{\phi}\pder{\psi}{r}\,\rmd\phi\,\rmd r.
\end{equation}
In fact, each of these integrals vanish. To see this, note that, since $\nabla^2\psi=0$, the derivatives inside the integrals may each be written as a Fourier series of the form
\begin{align}
\pder{\psi}{r} = \sum_mf_m(r,z)\Big(A_m\sin(m\phi) + B_m\cos(m\phi)\Big).
\end{align}
By orthogonality of the trigonometric functions, products for different $m$ vanish and we are left with integrals of the form
\begin{equation}
\sum_m\int_{-L}^L mg_m(r,z)\left(\int_0^{2\pi}\Big(A_m\sin(m\phi) + B_m\cos(m\phi) \Big)\Big(A_m\cos(m\phi) - B_m\sin(m\phi) \Big) \,\rmd\phi\right)\,\rmd z
\end{equation}
(and similar with $z$ replaced by $r$). But these integrals also vanish when integrated between 0 and $2\pi$. Therefore $H_{{\bf B}^v}({\bf B}^{\rm p})=0$, and it follows that $\Hw({\bf B}^{\rm p})=0$ for a periodic field in a cylindrical domain.

\subsubsection{Aperiodic Boundary Conditions} \label{sec:aper}

\begin{figure}[tb]
\begin{center}
\includegraphics[width=\textwidth]{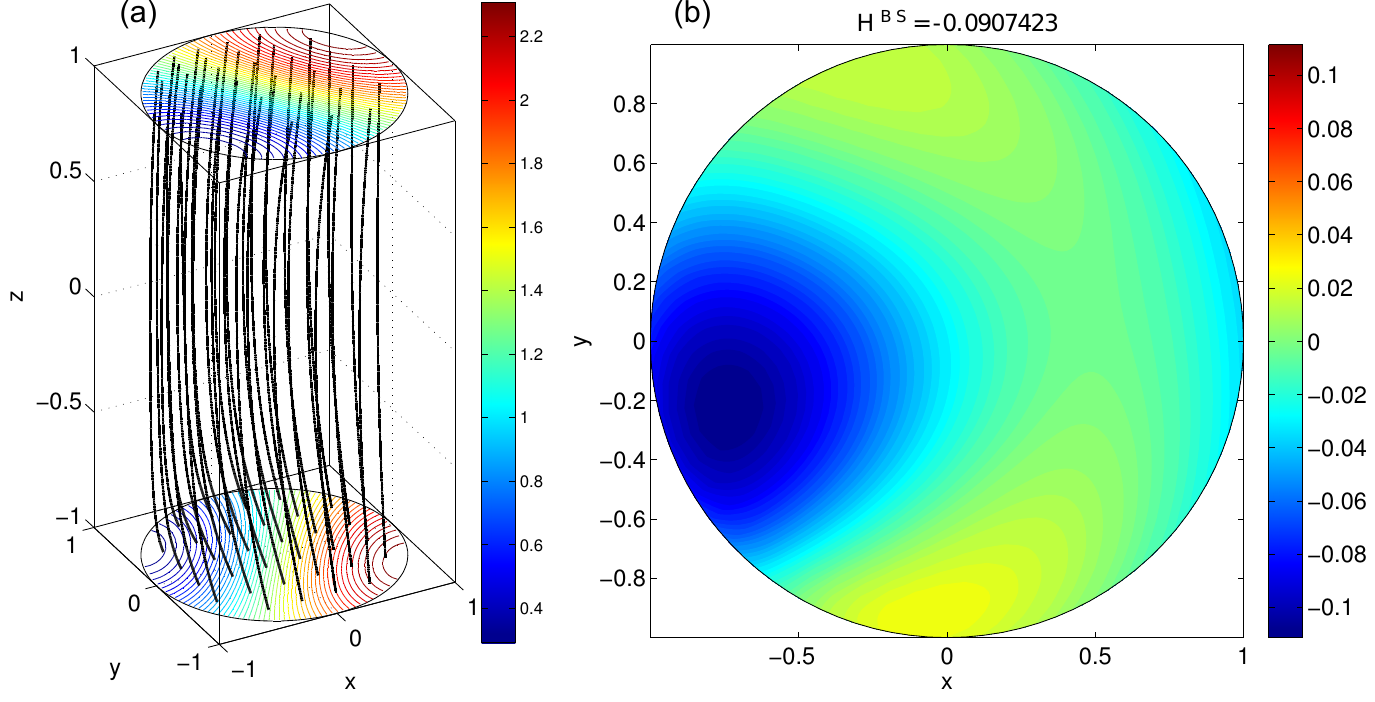}
\caption{A potential field with $\Hw\neq 0$, as described in Section \ref{sec:aper}. Panel (a) shows magnetic field lines and contours of $B_z$ on $S_0$ and $S_h$. Panel (b) shows contours of ${\cal A}^{\rm W}$ on $S_0$ (color shading), which integrate (weighted by $B_z$) to give a negative net $\Hw$.}
\label{fig:jlpf}
\end{center}
\end{figure}

If the magnetic field is aperiodic, i.e., $B_z(r,\phi,L)\neq B_z(r,\phi,-L)$, then the proof in Section \ref{sec:periodic} fails because there is no longer a straight, vertical magnetic field that matches the boundary conditions. In fact, for an aperiodic potential field, one might expect that the differing boundary conditions on $S_0$ and $S_h$ could introduce a ``twist'', in the sense of a net winding measured with respect to a fixed frame. To show that this is indeed the case, we present a specific example, studied previously by \citet{Janse2009b} and \citet{Low2011h}. Let ${\bf B}^{\rm p}$ be the specific field defined by the potential
\begin{equation}
\psi(r,\phi,z) = \frac{J_1(k_0r)}{k_0^2}\left(\frac{\sinh(k_0z)}{\sinh(k_0L)}\sin\phi + \frac{3\cosh(k_0z)}{2\cosh(k_0L)}\cos\phi \right),
\end{equation}
as in Equation \eqref{eqn:pot}, and let $R_0=L=1$. Here $J_1$ is a Bessel function of the first kind, and the constant $k_0$ must be chosen so that $B_r|_{S_s}=0$, which requires that $R_0k_0J_0(k_0R_0) - J_1(k_0R_0)=0$. We take the smallest solution $k_0\approx 1.8412$. This results in different distributions of $B_z$ on $S_0$ and $S_h$, namely
\begin{align}
B_z(r,\phi,-L) &= 1.3 + J_1(k_0r)\big(1.5\cos\phi - \sin\phi\big),\label{eqn:bz0}\\
B_z(r,\phi,L) &= 1.3 + J_1(k_0r)\big(1.5\cos\phi + \sin\phi\big). \label{eqn:bzh}
\end{align}
These are shown in Figure \ref{fig:jlpf}(a), along with a selection of field lines. The asymmetry between the two boundaries introduces a visible ``twist'' into the overall field, despite the fact that $\nabla\times{\bf B}^{\rm p}\equiv 0$. We have calculated the flux function ${\cal A}^{\rm W}$ for this field using Equation \eqref{eqn:fluxfun}, by numerically computing the pairwise winding numbers between a sample of field lines. This is shown in Figure \ref{fig:jlpf}(b). Integrating this flux function over $S_0$, weighted by $B_z(r,\phi,-L)$, we find that $\Hw\approx -0.09$. For this particular example, our numerical $\Hw$ converges to the same value as the cylindrical helicity defined by \citet{Low2011h}, which for this field is
\begin{equation}
H = -\frac{6\pi LJ_1^2(k_0R_0)}{k_0^2\sinh(k_0L)}.
\end{equation}
However, as we show in Appendix \ref{app:low}, the gauge ${\bf A}^{\rm CK}$ used by Low differs, in general, from ${\bf A}^{\rm W}$. Therefore they are measuring winding with respect to different frames. 

In summary, this example shows that an aperiodic potential field may have non-zero $\Hw$. In Section \ref{sec:equiv}, we will see one way of constructing an alternative field that has $\Hw=0$.

\subsubsection{Twisted Domains} \label{sec:coil}

\begin{figure}[tb]
\begin{center}
\includegraphics[width=0.5\textwidth]{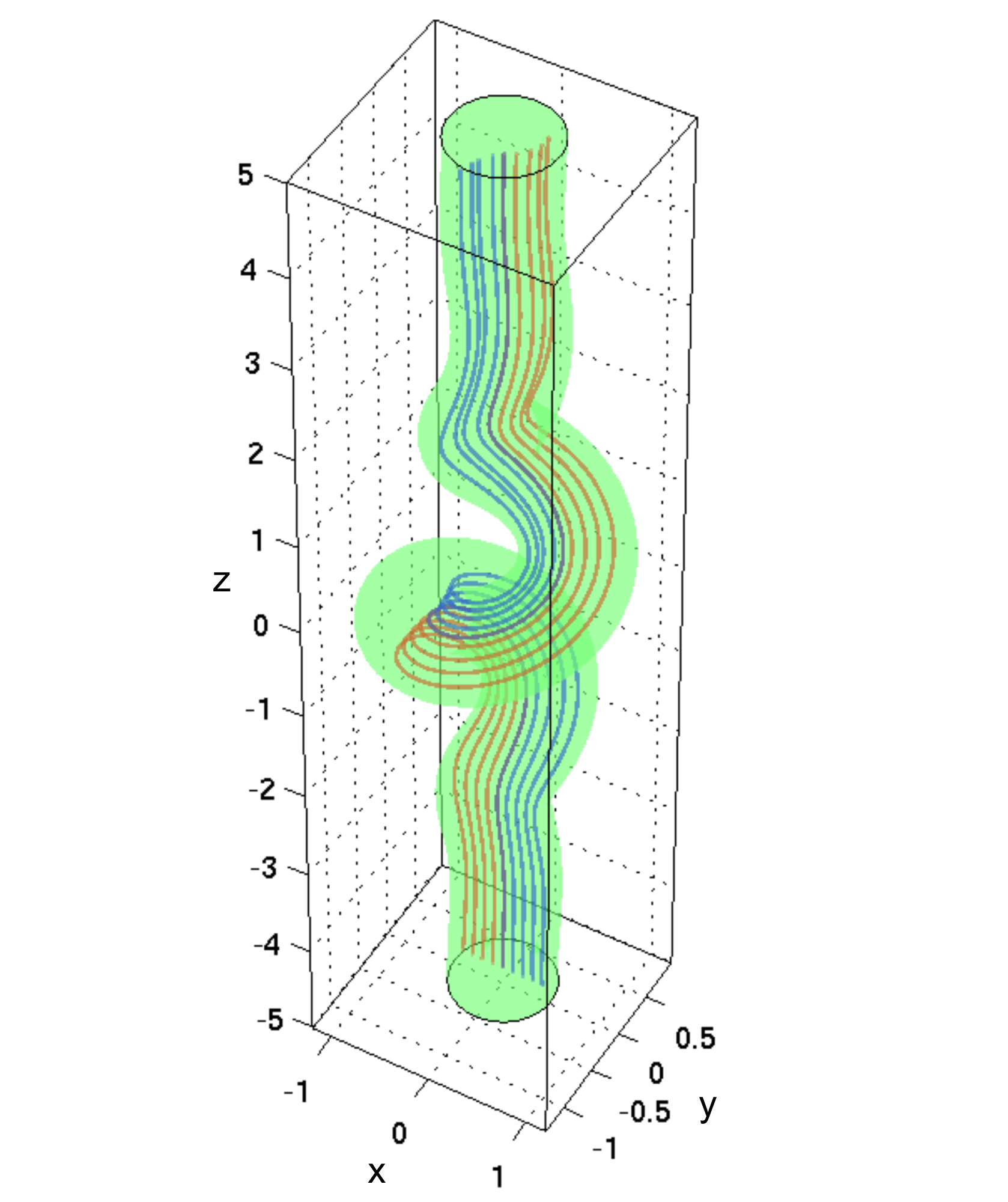}
\caption{A potential field in a coiled domain, calculated using a finite-element method. 
Selected field lines show that this potential field is uniformly twisted, at a rate corresponding to the writhe of the axis curve. Consequently it has non-zero winding helicity $H^{\rm W}({\bf B})\approx 0.16$.}
\label{fig:coil}
\end{center}
\end{figure}

Even if the boundary conditions are periodic, the potential field may inherit winding due to the shape of the domain. Figure \ref{fig:coil} shows an example where the potential field has non-zero $H^{\rm W}$ owing to the coiled shape of the domain, despite the fact that we have uniform boundary conditions $B_z=1$ on both $S_0$ and $S_h$. Up to an integer, ${\cal L}({\bf x},{\bf y})$ is the difference $\Theta({\bf x},{\bf y})(h) - \Theta({\bf x},{\bf y})(0)$, and it may be seen from the field lines plotted in  Figure \ref{fig:coil} that the field is uniformly-twisted (\emph{i.e.}, ${\cal L}({\bf x},{\bf y})$ is the same for all pairs of field lines). As this domain is tubular, we can decompose $\Hw$  into the sum of twisting ${\cal T}$  (the rotation of field lines about the central axis of the tube), and writhing ${\cal W}$ (a quantity measuring the self-winding of the tube's axis) 
 - see \cite{Berger2006}. Here we have confirmed numerically that the helicity $\Hw$ of the domain is equal to the writhe ${\cal W}$, and hence ${\cal T} =0$, meaning that the field has no internal twist about its axis. The writhing is a property of the axis shape alone, so in this case the potential-field helicity $\Hw$ is entirely determined by the shape of the domain. In general, aperiodic boundary conditions and/or non-potential fields on such domains will also have internal helicity from the twisting and braiding of field lines along the tube's length. It is possible for a field to have $\Hw=0$ on such a domain, but in this case it would require some internal twisting, unlike the cylindrical domain.

\subsection{Equivalence of relative helicity and standard helicity} \label{sec:equiv}

In what follows we once again restrict ourselves to a circular cylinder $V=\{0\leq r\leq R_0, \, 0\leq\phi < 2\pi, \, -L\leq z\leq L \}$ to provide clarity to the arguments. To substantiate our claim that the standard magnetic helicity and the relative helicity have equal physical meaning, we can use Equation \eqref{eqn:rbs} to prove that the gauge choice in $H({\bf B})$ and the choice of ${\bf B}'$ in $H_{{\bf B}'}({\bf B})$ are equivalent. Our result may be formulated as follows.
\begin{enumerate}
\item {\it Given any reference field ${\bf B}'$, we can always find a gauge in which $H({\bf B})=H_{{\bf B}'}({\bf B})$.}
\item {\it Conversely, given an arbitrary gauge for $H({\bf B})$, we can always find ${\bf B}'$ such that $H_{{\bf B}'}({\bf B})=H({\bf B})$.}
\end{enumerate} 
To prove part 1, note that in any gauge we can write
\begin{equation}
H({\bf B}) = \Hw({\bf B}) + \int_{S_h}\chi B_z\,d^2x - \int_{S_0}\chi B_z\,d^2x,
\end{equation}
where ${\bf A}={\bf A}^{\rm W} + \nabla\chi$. From Equation \eqref{eqn:rbs} we know that $H_{{\bf B}'}({\bf B})=\Hw({\bf B}) - \Hw({\bf B}')$, so we can simply choose the gauge to be
\begin{equation}
\chi = -\left(\frac{z+L}{2L\Phi_0}\right)\Hw({\bf B}'),
\end{equation}
which is a function of $z$ only. To prove part 2, note that we can give $H({\bf B})$ an arbitrary real value by changing gauge. Applying Equation \eqref{eqn:rbs} again, we must then show the existence of a reference field ${\bf B}'$ such that $\Hw({\bf B}')$ takes any arbitrary value. We show one way to explicitly construct such a reference field in Appendix \ref{app:ref}.

\begin{figure}[ht]
\begin{center}
\includegraphics[width=0.8\textwidth]{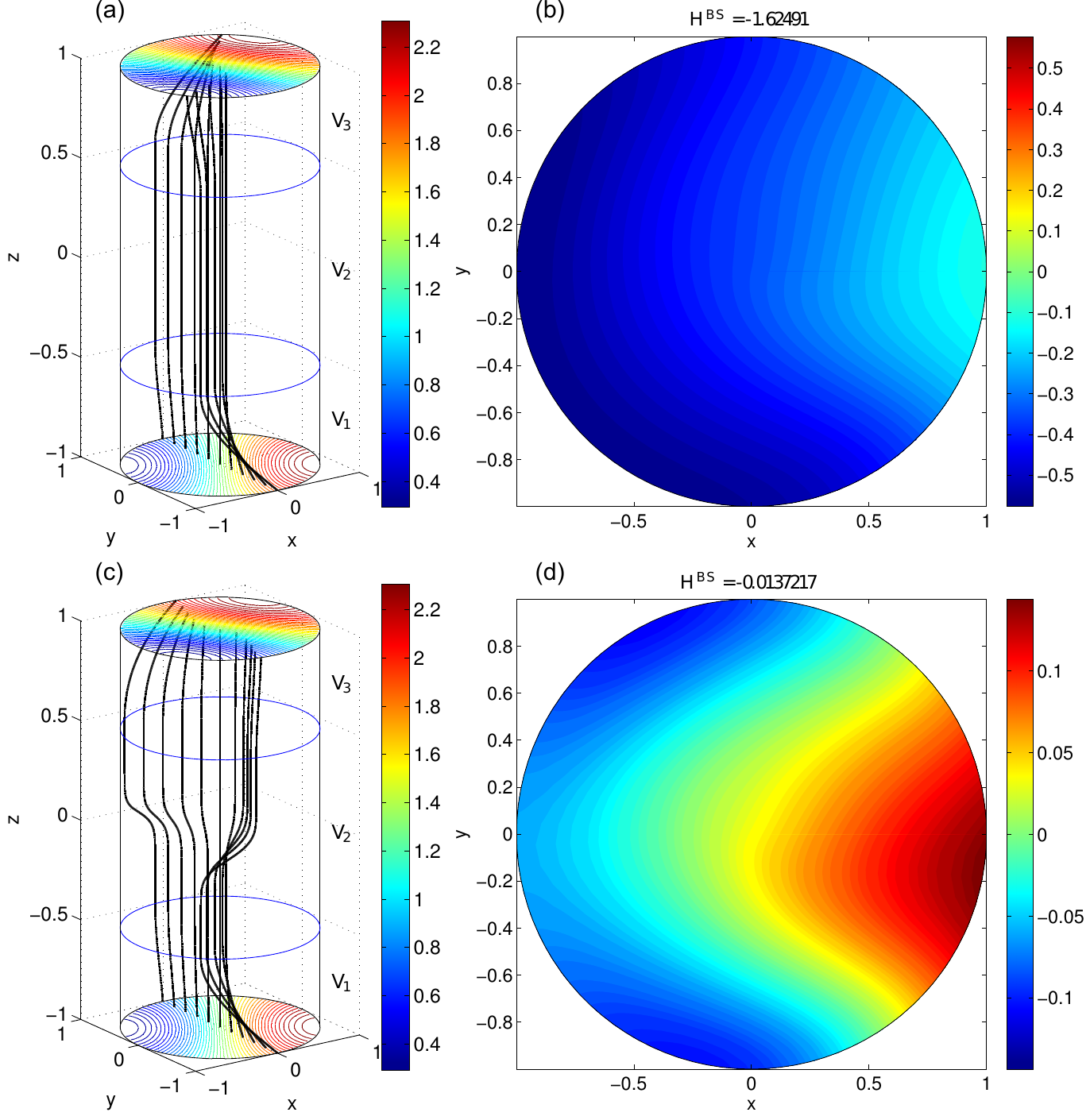}
\caption{A magnetic field with $\Hw=0$, constructed according to the method in Appendix \ref{app:ref}. Panels (a), (b) show the magnetic field and ${\cal A}^{\rm W}$ for $k=0$, while panels (c), (d) show corresponding plots for a non-zero value of $k$, chosen so that $\Hw\approx0$. Here we took $R_0=L=1$, $z_1=-0.5$, $z_2=0.5$.}
\label{fig:ref}
\end{center}
\end{figure}

One application of the technique in Appendix \ref{app:ref} is to construct a reference field with vanishing $\Hw$. To illustrate such a construction, let $V$ be the same circular cylinder as in Section \ref{sec:untwist}, and take the same boundary conditions \eqref{eqn:bz0}, \eqref{eqn:bzh} as for the potential field in Section \ref{sec:aper}. That potential field was ``twisted'', with $\Hw\neq 0$. By choosing the arbitrary constant $k$ appropriately in the field ${\bf B}_k$ (see Appendix \ref{app:ref}), we can construct an alternative magnetic field with $\Hw=0$. For this example, we have chosen $z_1=-0.5$, $z_2=0.5$, and $f(z)=\exp(-100z^2)$. The constructed fields with $k=0$ and with the (roughly) optimum value of $k\approx 4.14$ are shown in Figure \ref{fig:ref}. This field is likely not the only possible field with $\Hw=0$ (even its topology, as captured by the ${\cal A}^{\rm W}$ distribution, is likely not unique). Nor is it likely to be a stable equilibrium. It is presented here simply to prove that a field with $\Hw=0$ exists for arbitrary boundary conditions on the cylinder. Note that while $\Hw=0$, indicating that the average pairwise winding of the field lines vanishes, it is clear that ${\cal A}^{\rm W}=0$, so that individual field lines do see a net winding.

\section{Conclusions} \label{sec:conclusions}

To summarize, we have shown that, for open magnetic fields between two parallel planar boundaries, the helicity $H$ has a physical meaning in any gauge: it is the average pairwise winding number of magnetic field lines with respect to some frame field. We have shown how this gauge freedom is equivalent to the freedom of choice of reference field in the commonly used relative helicity for such fields. Moreover, there is a unique choice of gauge that always measures winding with respect to a fixed frame. This is the ``winding'' gauge of Equation \eqref{eqn:bs2}. We propose that the helicity in this gauge ($\Hw$) is a physically-motivated measure of the topological linking in an open magnetic field, that is uniquely defined and does not depend on choice of an arbitrary reference field. In effect, it always measures winding with respect to a straight field. As we have demonstrated in Section \ref{sec:gen}, from a geometrical perspective any other choice is unnecessary as it adds a contribution to the helicity arising form the rotation of the reference frame used to measure winding. This quantity has no physical meaning. Using the relative helicity with a potential reference field may or may not measure the same helicity $\Hw$, depending on the boundary conditions. If it differs from $\Hw$ then the relative helicity measure necessarily includes  a contribution due to a rotating reference frame.

As a result, we make the following practical recommendation. In magnetic fields having more than one boundary with $B_n\neq 0$, one should calculate $\Hw$, rather than using the relative helicity with potential field as a reference. For a cylindrical domain whose end boundaries $S_0$ and $S_h$ are the same shape, and if the boundary conditions on $B_n$ are the same on both ends ({\it i.e.} are periodic), then the two helicities are equal. This is because the potential field itself then has vanishing $\Hw$ in such a field, meaning that it is untwisted with respect to a straight field. But if these conditions are not met then the potential field will generally have $\Hw\neq 0$, so that the relative helicity does not match $\Hw$. For example, we have shown that this may arise if the boundary conditions are aperiodic (Section \ref{sec:aper}) or if the domain boundary has a complex shape (\ref{sec:coil}). We envisage that the absolute measure provided by $\Hw$ will be particularly useful when analyzing the time evolution of magnetic configurations where $B_n$ on the boundary is changing, or when comparing different magnetic fields. An example of the latter would be the comparison of different magnetic active regions in the solar corona \citep{Valori2012b}. 

In practical terms, there are several ways to calculate $\Hw({\bf B})$ for a given magnetic field ${\bf B}$. Using \eqref{eqn:rbs}, one can calculate the relative helicity with respect to a reference field known to have vanishing $\Hw$ (for example, Appendix \ref{app:ref} shows how to construct such a reference in the cylinder). Or, one can evaluate the vector potential by numerically evaluating \eqref{eqn:bs2}, then computing $\int_V{\bf A}^{\rm W}\cdot{\bf B}\,d^3x$. But the most straightforward method will generally be to utilize the physical interpretation and calculate $\Hw$ directly from ${\bf B}$ using \eqref{eqn:hbs}, evaluating pairwise winding numbers of field lines. We have implemented this numerically for the examples in Figures \ref{fig:jlpf} and \ref{fig:ref}.

It is interesting to observe that there are some similarities between the winding gauge choice and the gauges chosen by \citet{Hornig2006} and by \citet{Low2011h} in their suggested ``universal'' or ``absolute'' helicities. \citet{Hornig2006} fixes $H$ with the gauge condition that $\nabla^\perp\cdot{\bf A}=0$ everywhere on the boundary $\partial V$ (here $\nabla^\perp$ denotes the component of the gradient tangential to the boundary). On $S_0$ and $S_h$, this condition is satisfied by ${\bf A}^{\rm W}$ (this is seen directly from Equation \ref{bsiden}), but it is not satisfied in general by ${\bf A}^{\rm W}$ on $S_s$. \citet{Low2011h} defines an absolute helicity $H$ (for $V$ a cylinder) by taking the gauge
\begin{equation}
{\bf A}^{\rm CK} = \nabla\times\psi\hat{\bf z} + \eta\hat{\bf z},
\label{eqn:ack}
\end{equation}
which corresponds to a Chandrasekhar-Kendall representation of ${\bf B}$ for functions $\psi$, $\eta$. Our ${\bf A}^{\rm W}$ may also be written in this form with
\begin{equation}
\psi = -\frac{1}{2\pi}\int_{S_z}B_z({\bf y})\log|{\bf r}|\,\rmd^2{y},\quad \eta = A^{\rm W}_z.
\end{equation}
However, Low applies specific boundary conditions to uniquely define $\psi$ and $\eta$, and these are not the same as for ${\bf A}^{\rm W}$ in general (see Appendix \ref{app:low}). Another gauge condition, suggested by \cite{Valori2012b}, is that $A_z=0$; this does not uniquely specify the gauge, and the resulting freedom is used to make $H_{{\bf B}^{\rm p}}\equiv H$. However, we show in Appendix \ref{app:low} that the winding gauge ${\bf A}^{\rm W}$ may have ${A}^{\rm W}_z\neq0$, so in general this measure differs from $\Hw$.  We conclude in general that these proposed gauges measure the field-line winding in a non-physical rotating frame, yielding (in general) a different helicity measure from $\Hw$.

We conclude with some remarks about the generality of our results. We have assumed that our domain $V$ is simply-connected and lies between two parallel planar boundaries $S_0$, $S_h$. Furthermore, these parallel boundaries are the only part of the boundary where we allow $B_n\neq 0$. These restrictions are necessary so that the winding gauge ${\bf A}^{\rm W}$ is well-defined. In making the restriction that $S_0$ and $S_h$ are planar and parallel, we are essentially identifying a distinguished direction ($\hat{\bf z}$) that is perpendicular to the cross-sections $S_z$ on which ${\bf A}^{\rm W}$ is defined. This distinguished direction is also needed for defining winding numbers ${\cal L}({\bf x},{\bf y})$, along with a choice of coordinate frame $\{\hat{\bf e}_1,\hat{\bf e}_2\}$ on each cross-section. Extending the definitions of ${\bf A}^{\rm W}$ or ${\cal L}$ to a domain with curved boundaries $S_0$, $S_h$ would require choosing a foliation of \emph{curved} cross-sectional surfaces throughout $V$, complicating the definition of what it means for two curves to have non-zero winding number. We hope to address these complications in future.

\acknowledgments

ARY was supported by STFC consortium grant ST/K001043/1 to the universities of Dundee and Durham, and CP by an Addison-Wheeler postdoctoral fellowship. We thank the referee for interesting suggestions that have improved the paper. Figure \ref{fig:coil} used $i$FEM \citep{chen2009ifem} and DistMesh \citep{Persson2004}.

\appendix

\section{Construction of a magnetic field with arbitrary $H^{{\rm W}}$.} \label{app:ref}

In this Appendix, we give one method for constructing a magnetic field ${\bf B}$ in a circular cylinder $V$ to match arbitrary normal distributions on $S_0$ and $S_h$ which have a non-zero net flux, such that $\Hw({\bf B})$ is an arbitrary real number. Denote the two normal distributions by $B_z|_{S_0}=g_0(x,y)$ and $B_z|_{S_h}=g_h(x,y)$. We assume that $B_r|_{S_s} = 0$, so conservation of flux requires that $\int_{S_0}g_0\,\rmd^2x=\int_{S_h}g_h\,\rmd^2x$. We begin by assuming that $B_z$ can only have one sign on both boundaries, that is to say $g_0(x,y)$ and   $g_h(x,y)$ are either both positive definite or both negative definite.  

The basic idea of our construction is to divide $V$ into three distinct subdomains $V_1=V\cap\{z \,| \,z\in [-L,z_1] \}$, $V_2=V\cap\{z \,| \,z\in [z_1,z_2] \}$, and $V_3=V\cap\{z \,| \,z\in [x_2,L] \}$. We utilise the property that $H^{{\rm W}}$ is additive in $z$, i.e.,
\begin{equation}
\Hw(V) = \Hw(V_1) + \Hw(V_2) + \Hw(V_3).
\end{equation}
This follows from the winding-number interpretation in Section \ref{sec:wind}. In $V_2$ we will choose a magnetic field whose winding helicity is known and can be controlled. In $V_1$ and $V_3$ we will show how to construct magnetic fields that map the boundary flux distributions (on $S_0$ and $S_h$) to uniform flux distributions on the intermediate surfaces $S_{z_1}$ and $S_{z_2}$, so as to match on to the chosen field in $V_2$. These two fields will contribute some fixed $\Hw(V_1) + \Hw(V_3)$ that depends only on $g_0$, $g_h$, and not on the choice of field in $V_2$. By choosing the field in $V_2$ appropriately, we will obtain any desired $\Hw(V_2)$ and hence any desired $\Hw(V)$. An example magnetic field computed with this method is shown in Figure \ref{fig:ref}.

{ \it 1. Volume $V_2$.} In this region, we shall set
\begin{equation}
{\bf B}(V_2)={\bf B}_0 + {\bf B}_k,
\end{equation}
where ${\bf B}_0=B_0\hat{\bf z}$ and ${\bf B}_k=kf(z)(-x_2\hat{\bf e}_1 + x_1\hat{\bf e}_2)$. Here $k$ is an arbitrary constant, $f(z)$ is an arbitrary function of $z$ (for now), and $(x_1, x_2, z)$ are Cartesian coordinates. This field corresponds to an overall twist that varies in $z$. Let ${\bf A}_0$ and ${\bf A}_k$ be vector potentials for ${\bf B}_0$ and ${\bf B}_k$ in the winding gauge. Then
\begin{equation}
\Hw(V_2)=\int_{V_2}({\bf A}_0 + {\bf A}_k)\cdot({\bf B}_0 + {\bf B}_k)\,\rmd^3x.
\end{equation}
From the definitions of ${\bf B}_0$, ${\bf B}_k$ and the winding gauge, we have immediately that ${\bf A}_0\cdot{\bf B}_0=0$ and ${\bf A}_k\cdot{\bf B}_k=0$. We also find that
\begin{align}
{\bf A}_0\cdot{\bf B}_k &= \frac{kB_0f(z)}{2\pi}\int_{S_z}\frac{x_2(x_2-y_2) + x_1(x_1-y_1)}{(x_1-y_1)^2 + (x_2-y_2)^2}\,\rmd^2y,\\
{\bf A}_k\cdot{\bf B}_0 &= \frac{kB_0f(z)}{2\pi}\int_{S_z}\frac{-y_2(x_2-y_2) - y_1(x_1-y_1)}{(x_1-y_1)^2 + (x_2-y_2)^2}\,\rmd^2y.
\end{align}
Therefore we get an explicit expression for the winding helicity
\begin{equation}
\Hw(V_2) = \frac{\pi kB_0R_0^4}{2}\int_{z_1}^{z_2}f(z)\,\rmd z.
\end{equation}
In particular, by varying $k$ we may obtain any real value for $\Hw(V_2)$. We shall choose the function $f(z)$ so that ${\bf B}_k$ matches smoothly to zero at $z=z_1$ and $z=z_2$.

{\it 2. Volumes $V_1$ and $V_3$.} It suffices to consider $V_1$ (a similar construction will work in $V_3$). Our strategy is to first prescribe $B_z$ to be some function $\lambda(r,\phi,z)$ that interpolates between $\lambda(r,\phi,-L)=g_0(r,\phi)$ and $\lambda(r,\phi,z_1)=B_1$ (constant), then find suitable $B_r$, $B_\phi$ such that $\nabla\cdot{\bf B}=0$. (We work in polar coordinates.) For this, it is convenient to write ${\bf B}=\lambda{\bf v}$, where ${\bf v}=v_r(r,\phi,z){\bf e}_r + v_\phi(r,\phi,z){\bf e}_\phi + \hat{\bf z}$. Then from $\nabla\cdot{\bf B}=0$ we get
\begin{equation}
\frac{1}{\lambda}\frac{\rmd\lambda}{\rmd z} = -\nabla\cdot{\bf v},
\end{equation}
where the derivative is taken along field lines $r(z)$, $\phi(z)$ \citep[cf.][]{Yeates2012}.
The chain rule gives
\begin{equation}
\pder{\ln\lambda}{r}v_r + \frac{1}{r}\pder{\ln\lambda}{\phi}v_\phi + \pder{\ln\lambda}{z} = -\frac{1}{r}\pder{}{r}(rv_r) - \frac{1}{r}\pder{v_\phi}{\phi}.
\label{eqn:ln1}
\end{equation}
Since we have one equation for the two unknowns $v_r$, $v_\phi$, there is some freedom remaining. In order to satisfy $B_r=0$ on the side boundary $S_s$, we shall simply choose $v_r\equiv 0$ throughout $V$. (This means that all field lines of our constructed field will lie on concentric cylinders.) In that case, \eqref{eqn:ln1} reduces to
\begin{equation}
\pder{\ln\lambda}{\phi}v_\phi + r \pder{\ln\lambda}{z} = - \pder{v_\phi}{\phi}.
\label{eqn:ln2}
\end{equation}
At fixed $r$, $z$, this is an ordinary differential equation, readily solved to find
\begin{equation}
v_\phi(r,\phi,z) = \frac{\lambda(r,\phi,z)}{\lambda(r,0,z)}\left(v_\phi(r,0,z) - \frac{1}{\lambda(r,0,z)}\int_0^\phi r\pder{\lambda}{z}\,\rmd\phi\right).
\label{eqn:vphi}
\end{equation}
For each $r$ and $z$, the value $v_\phi(r,0,z)$ is an arbitrary constant; we may set all of these to zero. Finally, we shall impose the additional requirement on $\lambda$ that $\partial\lambda/\partial z|_{S_{z_1}}=0$. This will ensure that $v_\phi=0$ on $S_{z_1}$, so that our magnetic field in $V_1$ continuously matches to that in $V_2$. Our construction is completed by finding a suitable interpolant $\lambda(r,\phi,z)$. For each $r$, $\phi$, we may choose $\lambda$ to be the unique quadratic function $Q(z)$ such that $Q(-L)=g_0(r,\phi)$, $Q(z_1)=B_1$, $Q'(z_1)=0$. This is
\begin{equation}
\lambda(r,\phi,z) = B_1 + \frac{(z_1 - z)^2}{(z_1+L)^2}\Big(g_0(r,\phi) - B_1\Big).
\end{equation} 
Finally we note that, since the value of the helicities on $\Hw(V_1)$ and $\Hw(V_3)$ have no dependence on $k$, we can simply choose $k$ such that $\Hw(V)$ takes any desired value.

We now relax our assumption on the sign of the functions $g_0$ and $g_h$, allowing both signs, although our construction will require that the net flux $\int_{S_0}g_0\,\rmd^2x$ is non-zero. The field representation $\lambda{\bf v}$ is not valid where the function $\lambda$ is zero and consequently our argument on the domains $V_1$ and $V_3$ for mapping the $B_z$ distributions from $g_0$ and $g_h$ to a constant on the planes $S_{z_1}$, $S_{z_2}$ breaks down. To remedy this, we observe that the argument has no essential dependence on the relative size of the three domains. Thus we shrink the domain $V_1\cup V_2 \cup V_3$ to allow for an additional domain at each end, \textit{i.e} $V_0=V\cap\{z \,| \,z\in [-L,z_{a}] \}$, $V_1=V\cap\{z \,| \,z\in [z_{a},z_1] \}$, $V_2=V\cap\{z \,| \,z\in [z_1,z_2] \}$, $V_3=V\cap\{z \,| \,z\in [z_2,z_{b}] \}$  and $V_4=V\cap\{z \,| \,z\in [z_{b},L] \}$. On the domain $V_0$ we set ${\bf B}$ to the unique potential field satisfying $B_r=0$ on $S_s$, $B_z=g_0$ on $S_0$, and $B_z=B_1$ (constant) on $S_{z_a}$. Similarly, on $V_4$ we set ${\bf B}$ to be the equivalent potential field with $B_z=B_1$ on $S_{z_b}$. The same construction as before may then be used on $V_1$, $V_2$, and $V_3$. Since
\begin{equation}
\Hw(V) = \sum_{i=0}^4 \Hw(V_i),
\end{equation}
and only $\Hw(V_2)$ depends on $k$, we can simply alter $k$ to obtain any desired value of $\Hw(V)$.

\section{Demonstrating ${\bf A}^{\rm CK} \neq {\bf A}^{\rm W}$ with Low's specification of $\psi$}\label{app:low}
The gauge choice used by \citet{Low2006f,Low2011h} in \eqref{eqn:ack} is such that the function $\eta\equiv A^{\rm CK}_z$ is zero on the boundary of the discs $S_z$. We demonstrate here that there is at least one admissible field on the cylinder for which ${A}^{\rm W}_z\neq 0$ at some point on the boundary $S_s$, so that ${\bf A}^{\rm CK}\neq{\bf A}^{\rm W}$. The field 
\begin{equation}
{\bf B} = -x_2^2{\bf \hat{e}}_1 + x_1x_2{\bf \hat{e}}_2 + (B_0 - x_1z){\bf \hat{z}}.
\end{equation}
is divergence free and tangent to the side boundary of the cylinder. At the particular point $(0,-R_0)$ on the side boundary, one may show by direct calculation that $A_z^{\rm W}(0,-R_0)=R_0^3/8$.
%

\bibliographystyle{apj}
\bibliography{yp}

\end{document}